\documentclass[floatfix,aps,prd,abstract,twocolumn,10pt,nofootinbib,preprintnumbers]{revtex4-1}

\usepackage{amsgen,amsmath,amstext,amsbsy,amsopn,amssymb}
\usepackage{graphicx}
\usepackage[usenames]{color}
\usepackage{epsfig}
\usepackage{multirow}
\usepackage{longtable}

\newcommand{\mb}{\mathbf}

\newcommand{\diff}{ \, \mathrm{d} }

\newcommand{\Ddel}{\delta_{\rm D}   }

\newcommand{\comment}[1]{}

\begin{document}


 \title{ Halo Sampling, Local Bias  and Loop Corrections }

\author{Kwan Chuen Chan} \email{kcc274@nyu.edu}
\author{Rom\'an Scoccimarro}

\affiliation{Center for Cosmology and Particle Physics, Department of Physics, \\
 New York University, NY 10003, New York, USA} 

\date{\today}

\begin{abstract}

We develop a new test of local bias, by constructing a locally biased halo density field from sampling the dark matter-halo distribution. Our test differs from conventional tests in that it preserves the full scatter in the bias relation and it does not rely on perturbation theory. We put forward that  bias parameters obtained using a smoothing scale $R$ can only be applied to computing the halo power spectrum at scales $k\sim 1/R$. Our calculations  can automatically include the running of bias parameters and give vanishingly small loop corrections at low-$k$. Our proposal results in much better agreement of the sampling and perturbation theory results with simulations.  In particular, unlike the standard interpretation of local bias in the literature, our treatment of local bias does not  generate a constant power in the low-$k$ limit. We search for extra noise in the Poisson corrected halo power spectrum at wavenumbers below its turn-over and find no evidence of significant positive noise (as predicted by the standard interpretation) while we find evidence of negative noise coming from halo exclusion for very massive halos. Using perturbation theory and our non-perturbative sampling technique we also demonstrate that nonlocal bias effects discovered recently in simulations impact the power spectrum only at the few percent level in the weakly nonlinear regime.

\end{abstract}

\maketitle

\section{Introduction}
Modern galaxy surveys are wide and deep enough to allow us to compute the statistics such as the power spectrum and bispectrum (or their real space counterparts).  However, galaxies are known to be biased objects relative to the underlying dark matter  \cite{Kaiser84}, thus we must understand the bias relation in order to extract information on the dark matter, which contains cosmological information.  As a first step towards understanding galaxy bias, we need to understand the bias relation for dark matter halos, where galaxies are thought to reside.  There exists  several prescriptions for the halo bias predicting the bias parameters from the initial conditions. These approaches assume some heuristic mechanisms for the identification of halos from the initial density field. The peak bias formalism \cite{Kaiser84, BBKS86, Desjacques08} assumes that galaxies form at the peaks of the initial density field.  Another popular model is the peak-background split, in which the density field is split into long and short wavelength parts. The long wavelength perturbation modifies the background density and the short wavelength causes local galaxy formation \cite{Kaiser84,BCEK1991,MoWhite, ShethLemson}. All these approaches are valuable for understanding the physics of bias,  when compared to simulations they work remarkably well, but are not accurate at percent level. For example, although it is found that about 70\% of the well-resolved halos can be associated with  peaks in the initial density field \cite{LudlowPorciani}, a significant fraction of halos that do not correspond to peaks.  Comparison of the linear bias from peak-background split calculations shows that there are still 5\% discrepancies with $N$-body simulations at large scales~\cite{Marc2007}.

A commonly adopted approach at large scales is to use the local biasing prescription \cite{FryGaztanaga}, \textit{i.e.}~the smoothed halo density field $\delta_{\mathrm{ h} R }  $ is assumed to be a local function of the smoothed dark matter density $\delta_{ R } $ 
\begin{equation}
\label{eq:EulerianLocalBiasing}
\delta_{\mathrm{ h} R } (\mb{x} ) =  \sum_{i=0}^{\infty}    \frac{b_i^R }{i !}    \big( \delta_{  R   }( \mb{x}  )  )^i ,
\end{equation}
where $b_i^R$ is a free parameter. We note that this local prescription only holds for the smoothed density fields, in which the small scale modes have been removed.  We should regard Eq.~(\ref{eq:EulerianLocalBiasing}) as a large-scale effective description.  In the bias parameters, we have used the superscript $R$ to emphasize that the Eulerian bias parameters in general depend on the smoothing scale. This is the prescription adopted in interpreting  modern survey data for two-point and in particular higher-order statistics \cite{GaztanagaFriemanAPM, ScoccimarroetalIRAS, FeldmanetalPSCz, Verdeetal2dFGRS, Gaztanagaeetal2dFGRS, McBrideeetalSDSS}.

As the local biasing prescription is the simplest well-motivated model, a number of works have been devoted to testing it \cite{MoWhite, Somervilleetal, Smith2007, ManeraGaztanaga2011, RothPorciani, Pollacketal}.  In practice, the series is often truncated at the quadratic order, sometimes at the cubic order.  When being tested against simulations, a quadratic or cubic polynomial is fitted to the scatter plot between $\delta_{R } $ and $\delta_{\mathrm{h} R  } $   to extract the bias parameters.  Also in the test perturbation theory is often invoked to compute the correlators, such as the halo power spectrum. There are two  peculiar features of the local biasing prescription that deserve further examination.  The bias parameters in the scatter between $\delta_R  $ and $ \delta_{\mathrm{ h} R}  $ vary when the smoothing scales changes \cite{ManeraGaztanaga2011, RothPorciani,Pollacketal}. On these grounds this method is sometimes dismissed as an unreliable way to measure the bias parameters, \textit{e.g.}~\cite{Pollacketal}.   In addition, unless the calculation is stopped at the tree level, it is well-known that higher-order loop corrections give rise to large undesirable corrections even at large scales unless an aggressive smoothing window is applied \cite{Heavens1998,McDonaldReBias}.  Both of these problems are especially obvious when the smoothing scale is small. For small smoothing scales, however, one may need to worry about the accuracy of fitting to the scatter plot  using a simple polynomial, as the scatter in the scatter plot is generally large (eventually signaling the breakdown of the local bias approximation). \comment{ For example \cite{RothPorciani} claimed that fitting by a cubic polynomial retains much more information than using a quadratic one. } One may also worry that the large corrections to the tree-level results may indicate the breakdown of the perturbation method.

In this paper, we develop a method, which does not suffer from these two potential issues. We perform a test without using an explicit parametrization, and so we effectively test local biasing without truncation. In fact, our test can keep all the information in the scatter plot. Furthermore, the correlators are computed numerically, and so we do not need to worry about the potential breakdown of perturbation theory. The clean test that we develop here is complementary to the conventional tests, and helps to highlight that there is a real problem to be addressed in the standard formulation of local bias.  We shall offer a new interpretation for computing the power spectrum using the local biasing prescription. Essentially we argue that the smoothing scales in the halo biasing prescriptions is a ``knob'' controlling the scales of the halo fluctuations that the biasing prescription measures.  The bias parameters obtained with large  smoothing window must be used to calculate large-scale predictions, and those with small smoothing scales should be used for small-scale statistics. This intuitive viewpoint does not seem to be widely appreciated (see however section~V in \cite{Scoccimarroetal2011}).  This interpretation naturally accommodates the two  ``undesirable'' effects of local biasing, the running of bias parameters in the scatter plot and the large loop corrections to tree-level results.

In~\cite{NLBias} we showed that under very general conditions that the local biasing prescription is not self-consistent, in the sense that even if galaxies are formed with local biasing initially, subsequent large scale gravitational evolution inevitably induces nonlocality (see also \cite{Fry96,Catelan1998}). The explicit form of the nonlocal terms up to third order were derived first assuming halo conservation, and then generalized to include a local source to take into account of the galaxy formation and merging. In both cases the structures of the nonlocal terms are the same. Ref.~\cite{NLBias} also measures the specific nonlocal terms in the scatter plot and bispectrum (see also \cite{Baldaufetal2012}). Other arguments also lead to the existence of extra contributions to the biasing prescription. Ref.~\cite{McDonaldRoy} proposes various correction terms using the invariants constructed from the physical fields. From Lagrangian perturbation theory, the halo bias is also known to be nonlocal \cite{Catelan1998,MatsubaraNLBias}. Here we develop a general test that enables us to investigate the importance of the nonlocality in a model-independent way. Because the effects of the extra nonlocal contributions is not large, it is vital to have a clean test. As we discussed above, our test is free of the the issues of standard tests, and this helps to pin down the effect of the extra contributions.  In this paper we focus on the halo power spectrum however, which turns out to be not effective in detecting the  nonlocality. In a forthcoming paper, we will extend the analysis to the bispectrum.

This paper is organized as follows. We describe the sampling technique in Section \ref{sec:sampling_technique} and the numerical results are presented in Section \ref{sec:numerical_results}. As a comparison and to help interpret the sampling results, in Section \ref{sec:PkFromPT} we present the results from standard Eulerian perturbation theory. We offer a new interpretation of Eulerian local biasing in Section \ref{sec:PkWithBiasRunning}, which results in much better agreement between sampling and simulation results.  As a case study, we study the power spectrum of a nonlocal model derived in \cite{NLBias} in section \ref{sec:NLPk}. In  section \ref{sec:NonlocalvsLocal} we investigate how the results would be biased when a local model is assumed if the underlying model is nonlocal. We summarize and conclude in Section \ref{sec:Conclusion}.  


\section{The Halo Sampling Method}
\label{sec:methodandresults}
In this section we first outline the halo sampling method and then present numerical results from it. 

\subsection{Description of the sampling technique}
\label{sec:sampling_technique}

In this section we describe a method to construct a halo density field via sampling the dark matter-halo distribution.  We shall abbreviate this method as \textit{halo sampling} or \textit{sampling}.  Here we outline the procedure.

To begin with, let us suppose that we have the dark matter particles and  halos from an $N$-body simulation.   First, we interpolate the dark matter field to a grid, and then Fourier transform it. The field is smoothed by applying some window function. In this paper we shall use a top-hat window with radius $R$ in real space throughout. After inverse Fourier transform,  the smoothed dark matter field is obtained.  The same procedure is carried out for the halos.  At each grid point, we have the dark matter and halo density field pair, $( \delta_{ R } ,   \delta_{\mathrm{h} R } )_i$, where the index $i$ denotes the coordinates of the grid. If we go on to make a scatter plot using these dark matter-halo pairs then this is the usual procedure to determine the bias parameters from the scatter plot. Instead, from these two smoothed fields,  their joint probability distribution $ f( \delta_{ R } ,   \delta_{ \mathrm{h} R }  ) $ is constructed. Note that the distribution $f$ is a one-point distribution only. With the joint distribution, for each grid point, given the dark matter field at that point $ \delta_{  R } $, we can construct a new halo field by drawing halo densities from the conditional probability distribution $ f(  \delta_{ \mathrm{h} R }  |  \delta_{  R }  ) $. The new halo field is constructed by sampling the dark matter-halo distribution, that is why we called this method  \textit{halo sampling}.

There is an easy way to carry out the sampling, and we shall dub it \textit{shuffling}. We first sort $( \delta_{ R } ,   \delta_{\mathrm{h} R } )_i$ in ascending order in  $\delta_{ R } $, and then bin them.  The $ \delta_{\mathrm{h} R } $  components of these pairs are permuted within the same bin in $\delta_R $. The shuffled halo field is put back to the grid using its grid index $i$. Thus each shuffling procedure is a realization of the conditional distribution $ f( \delta_{ \mathrm{h} R } | \delta_{  R }  ) $.

 Shuffling leaves the halo field unchanged if the halo field is solely determined by the dark matter density field. This shuffling procedure also leaves the $\delta_{\mathrm{h} R  }-\delta_R$ scatter plot unchanged.  Thus we can test the local biasing prescription by reproducing the same scatter. If the biasing prescription has extra degrees of freedom other than $\delta_R $ (and the stochastic noise \cite{DekelLahav}), the biasing prescription is called \textit{nonlocal}. The presence of nonlocality will cause deviation between the power spectrum of the original halo field and that of the sampled field.

  Suppose that the halo density is given by  $\delta_{\mathrm{h} R } = g( \delta_{  R }, \zeta )$, where $\zeta $ collectively denotes the extra degrees of freedom, including the stochastic noise. When we construct the scatter plot between $\delta_{\mathrm{h} R }$ and $\delta_{ R}$, we project the function onto the $\zeta=0$ hypersurface. After projection, some information in the original relation is lost. Using the projected distribution to compute the statistics will typically lead to a biased answer.   Because in our case we keep the information of the whole scatter plot, and we compute the correlators numerically, the shuffling results are probably the best that local biasing can ever achieve.

As a further simplification, instead of using the whole distribution, we shall also replace $ \delta_{\mathrm{h} R}$ in each $ \delta_{ R } $ bin by its mean. This simply reduces to the usual mean biasing prescription. We shall refer to this method as \textit{mean}.  Since we do not assume a specific functional form for the mean bias relation we effectively test Eq.~(\ref{eq:EulerianLocalBiasing}) without truncation. However, it turns out that the differences between the shuffling results and those using the  mean are negligible because the shuffling operation also destroys any small-scale stochastic noise, including the Poisson shot noise.  See Appendix \ref{sec:ShufflingStochasticNoise} for more details.


\subsection{ Numerical results of sampling}
\label{sec:numerical_results}

Before presenting the numerical results, we shall give some details about the simulations used in this work. The $N$-body simulations consist of $1280^3$ particles, in a box of size 2400 Mpc/$h$, which is part of LasDamas project \cite{LasDamas}. The cosmology is a flat $ \Lambda  $CDM model with $\Omega_{\rm m}  =0.25 $, $\Omega_{\rm \Lambda}  =0.75 $, and $\sigma_8 = 0.8$.  Thus the particle mass is $ 4.57  \times 10^{11} \, M_{\odot}/h$. The initial conditions are Gaussian, with a scale-invariant spectrum. The transfer function is output from CMBFAST \cite{CMBFAST}.  The initial particle displacements are implemented using second order Lagrangian perturbation theory \cite{2lpt} at redshift 49. The simulations are done using the $N$-body code Gadget2 \cite{Gadget2}.  We shall average over 10 realizations.

The halos are identified using the FOF algorithm \cite{Davisetal1985} with linking length of 0.156 of the mean interparticle separation. We consider halos with at least 20 particles. We shall focus on halos at two redshifts, 0.97 and 0 respectively.  At each redshift we further divide them into two  groups, adjusting the boundary so that each group has the same number density, and hence similar shot noise properties. The mass and number density of the halo groups are shown in Table \ref{tab:HaloGroup}.

\begin{table}
\caption{ Properties of halo groups used in this paper  }
\label{tab:HaloGroup}
\centering
\begin{ruledtabular}
\begin{tabular}{|l|l|l|l|l|}
  halo   &  redshift  &  mass range                    &   mean $\nu  $       &  number density                 \\ 
  group &            &    $(10^{12} \, M_{\odot}/h )$ &                               & $ (\mathrm{Mpc}/h )^{-3} $      \\   
\hline  \hline
I             &  0.97      & $ 9.15 - 14.2 $             &      1.97                    & $ 8.6 \times  10^{-5}   $       \\ \hline
II            &   0.97     & greater than  14.2         &       2.30                   & $ 8.6 \times  10^{-5}   $       \\ \hline
III           &   0        &  $9.15 - 16.5 $               &       1.29                   & $ 2.0 \times  10^{-4}   $       \\ \hline
IV            &   0        & greater than 16.5           &       1.61                   & $ 2.0 \times  10^{-4}   $       \\ 
\end{tabular}
\end{ruledtabular}
\end{table}

The dark matter particles and the halos are separately interpolated to grids of size $800^3$ using a fourth-order interpolation kernel. After FFT, we apply a top-hat window to smooth both density fields. Then we inverse FFT to real space to get the density pair $(\delta_{ R}, \delta_{\mathrm{h}R}  )$. We use shuffling as the implementation of the sampling technique. Thus we will sort the dark matter-halo pairs using the value of the dark matter density. After sorting, we then bin the pairs into 40 bins, which are of equal spacing in $\delta_{R}$. The $\delta_{\mathrm{h}R }$ components in each bin are then randomly permuted. When we implement the mean biasing instead of shuffling, the  $\delta_{\mathrm{h}R }$ components are replaced by their mean in that bin.  Subsequently, the modified halo density field is put back to the grid, and the power spectrum is computed numerically.

We now compare the power spectrum from the original field without smoothing with the one obtained from sampling the dark matter-halo distribution. For the most part we shall focus on the cross halo power spectrum to avoid dealing with shot noise subtraction, which may not be precisely Poisson. We will comment more on this in section~\ref{sec:PkWithBiasRunning}.

\begin{figure*}[!t]
\centering
\includegraphics[ width=\linewidth]{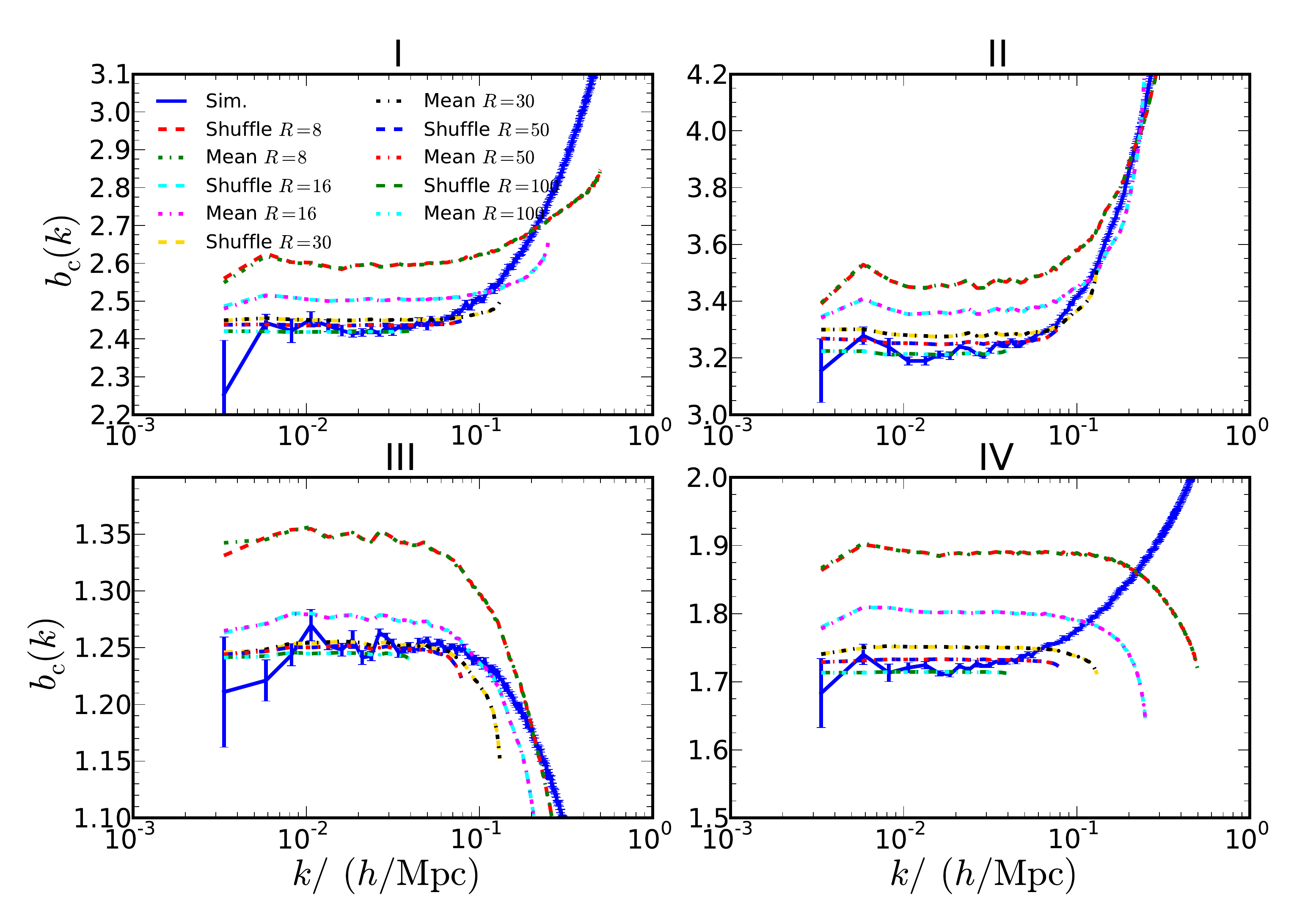}
\caption{  The bias $ b_{\rm c} $ derived from the cross power spectrum obtained from the original simulation (lines with error bars) and those sampled by the shuffling method (dashed) and the mean method (dot-dashed). The panels correspond to the four halo groups used in this paper, whose properties are given in Table \ref{tab:HaloGroup}. Although five different smoothing scales are used, the window function in the cross power spectrum and the dark matter cancel out neatly in the case of the original simulation. The error bar is derived from the standard deviation of the cross power spectrum using 10 realizations. The results from shuffling almost completely overlap with those using the mean for each smoothing scale. The five smoothing scales are 8  16, 30, 50 and 100 Mpc/$h$ (from top to bottom at large scales in each subplot). For each curve, we show up to the scale at which it starts to rapidly oscillate due to the window function. 
The discrepancy at large scales will be addressed in detail in Section  \ref{sec:PkWithBiasRunning}, it does disappear when we properly interpret local bias: small smoothing scales are only relevant for the high-$k$ power spectrum. 
}
\label{fig:bc_Sim_Shuffle_Mean_800} 
\end{figure*}

 In Fig.~\ref{fig:bc_Sim_Shuffle_Mean_800} we plot the bias parameter $b_{\rm c} $ derived from the cross power spectrum between halo and dark matter $P_{\rm c}  $ and the dark matter power spectrum $P_{\rm m }  $
\begin{equation}
b_{\rm c}  \equiv  \frac{ P_{\rm c}}{P_{\rm m }}.
\end{equation}
For the $b_{\rm c}$ from simulations, the effects of the smoothing window cancel out neatly.  This is because after multiplying the window functions, the smoothed fields are used to compute the power spectra directly. Therefore $b_{\rm c}$ obtained using different smoothing scales is identical, and we only show one curve. We emphasize that changes in $b_{\rm c} $ are equivalent to those in the halo power spectra since the dark matter field is not changed by the sampling procedure. 

 For small smoothing scale $R$, the $b_{\rm c}$ from sampling is larger than that from the original simulation even at large scales. As the smoothing scale increases, the sampling result approaches the original simulation result at large scales. Among the four groups shown here, the sampled power spectrum is always larger than the original one when $R$ is small. The sampled cross power spectrum agrees with the original one only when $R$ is as large as 50 Mpc/$h$. Except for the overall overshoot, $b_{\rm c}$ from the sampled field using small smoothing scales is scale independent up to around $ k \sim 0.1 \, h / \mathrm{Mpc} $. We also note that the two sampling techniques result in almost identical cross power spectra. The extra random noise in the shuffled field is averaged out when the halo field is crossed with the dark matter.

The rest of the paper is mainly devoted to understanding the physical significance of the smoothing scales in the local biasing prescription, and in particular why there is such a discrepancy in the power spectrum at large scales. We will elaborate it in detail in Section \ref{sec:PkWithBiasRunning}. However, the bottom line is rather intuitive. For each smoothing scale $R$ the halo biasing prescription measures the fluctuations of the halo density field at that particular scale.  The local biasing prescription with a large $R$ characterizes the fluctuations of the halo density field at large scales, and so we are only entitled to compute the halo power spectrum using that prescription for small $k$ in the power spectrum. Similarly, a local biasing prescription obtained with a small $R$ can only be used to compute the power spectrum at large $k$.  

\begin{figure*}[!t]
\centering
\includegraphics[ width=\linewidth]{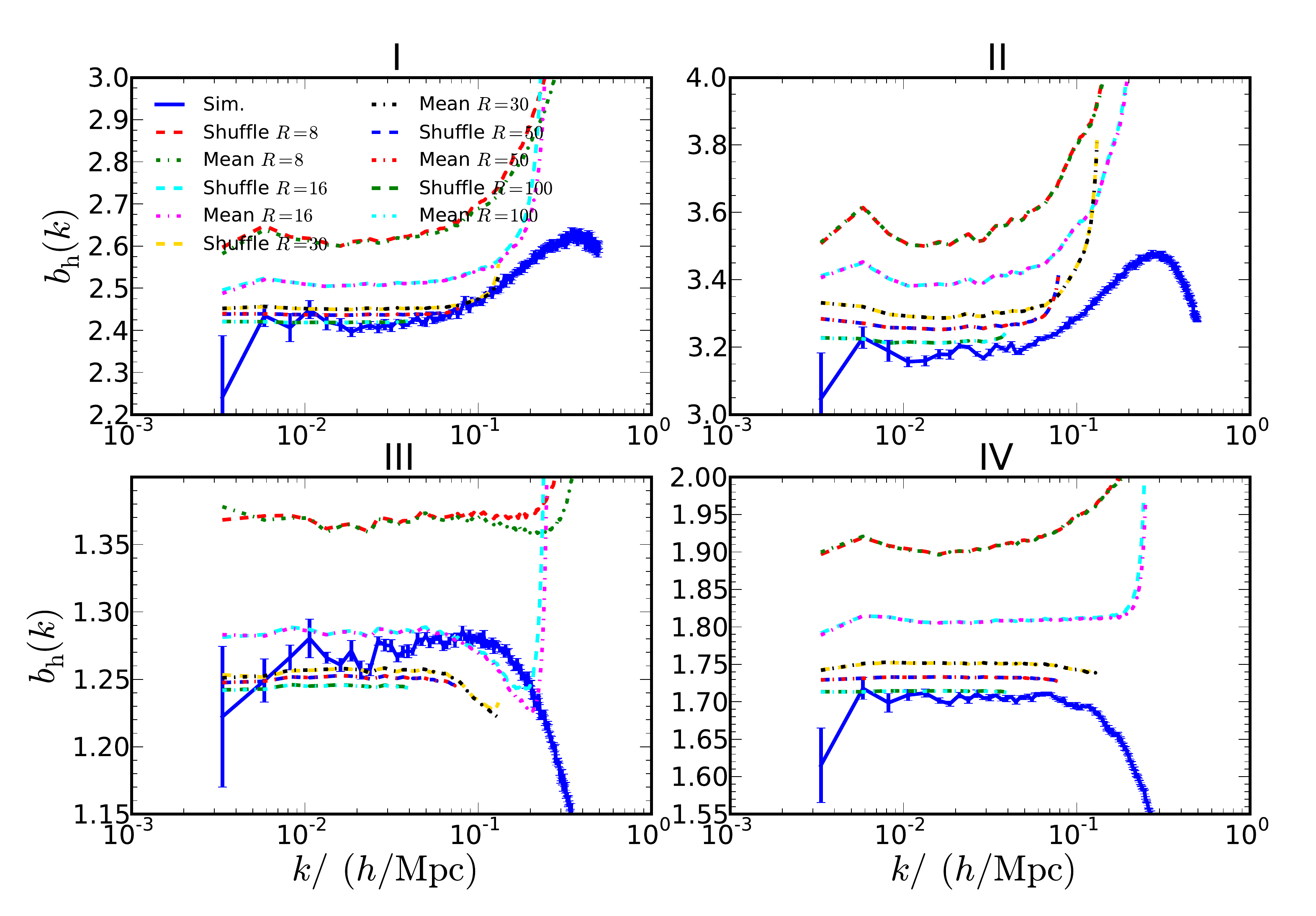} 
\caption{ Same as Fig.~\ref{fig:bc_Sim_Shuffle_Mean_800} except for the bias $b_{\rm h}$ computed using the halo power spectrum. For the simulation halo power spectrum, shot noise is subtracted assuming Poisson statistics, while there is no shot noise subtraction for the sampled halos (see main text as to why this is a reasonable thing to do).      }
\label{fig:bh_Sim_Shuffle_Mean_800} 
\end{figure*} 
First of all, we need to discuss the treatment of the shot noise. For the simulated halo power spectrum, we subtract the smoothed shot noise, which we assume to be  the usual Poisson shot noise \cite{Peebles} except with the window function
\begin{equation}
\label{eq:Pkshot}
P_{\mathrm{shot} R}(k)  \equiv   \frac{1  }{ (2 \pi)^3 } \frac{1}{\bar{n}_0 } W^2(kR), 
\end{equation}
where $ \bar{n}_0 $ is the mean number density of the halos. See Appendix \ref{sec:PoissonShotNoise} for its derivation. For the halo groups used in this paper, their number density is given in Table \ref{tab:HaloGroup}.

We now turn to the halo power spectrum. In Fig.~\ref{fig:bh_Sim_Shuffle_Mean_800}, where we show the bias derived from the halo (auto) power spectrum, that is  
\begin{equation}
b_{\rm h} \equiv \sqrt{ \frac{P_{\rm h} }{  P_{ \mathrm{m} } }}.    
\end{equation}

The shot noise is significant at around $k \sim 0.1 - 0.2   \, h/\mathrm{Mpc}$, at which the amount of noise subtracted becomes comparable to the signal (see Fig.~\ref{fig:Ph_Pshot}). Furthermore, the shot noise may be sub-Poisson due to halo exclusion effects, which is relatively more important for massive halos \cite{Smith2007}. There may be other types of stochastic noise as well. Complications in treating the shot noise  will prevent us from comparing the theory with simulations in a precise way at $k \gg 0.1  \, h/\mathrm{Mpc}$ for the halo power spectrum. Most of the work in this paper deals with the cross power spectrum for this reason. However we will come back to the halo power spectrum in section~\ref{sec:PkWithBiasRunning}, although we will concentrate there in the low-$k$ limit where shot noise issues are less of a concern.

As mentioned, due to discreteness, the halo density field directly extracted from simulations has shot noise. If we construct the sampled halo field by drawing from a continuous dark matter-halo distribution or simply using the mean biasing relation, the correlation of the sampled field is endowed from the dark matter field, and thus it does not have shot noise. However, it is not so obvious that the halo density field loses its shot noise after shuffling. The shot noise originates from the discrete nature of the halos. Heuristically, after smoothing, the halos spread out and shuffling further distributes the halo clouds everywhere, and so the correlations due to the discrete nature of halos is destroyed. In fact, any stochastic noise (not correlated with $\delta$) is in general is destroyed by the shuffling process (we have verified this experimentally by generating random particles and performing the shuffling process to the corresponding density field). See Appendix \ref{sec:ShufflingStochasticNoise} for more details. Thus we do not subtract shot noise from the sampled halo power spectrum in Fig.~\ref{fig:bh_Sim_Shuffle_Mean_800}, while we do subtract Poisson shot noise using Eq.~(\ref{eq:Pkshot}) from the numerical power spectrum (where no shuffling has been done).

However, if the Poisson shot noise is not accurate enough or there are additional stochastic noises there will be discrepancy between the sampling results and the simulation results.  Indeed, even for smoothing scales of $100 \, \mathrm{Mpc}/h $, there are some hints of percent-level deviations between the original power spectrum and the sampled one at large scales for halo group III (see bottom left panel in Fig.~\ref{fig:bh_Sim_Shuffle_Mean_800}). It is possible to interpret these  deviations as a small extra white noise contribution (giving a total super-Poisson shot noise), as opposed to a slightly different bias, since the matter power spectrum does not vary that much over the range of scales $k < 0.03 h/\mathrm{Mpc}$. From the cross spectrum in Fig.~\ref{fig:bc_Sim_Shuffle_Mean_800}, we also see  better agreement at large scales between the sampling and the original simulation results. However,  note that these halos are not well resolved by our simulations.  On the other hand, for groups II and IV the simulation results for the halo spectrum are slightly higher than those from sampling at large scales (an effect that is again not seen in the cross-power spectrum results). This might be due to small sub-Poisson contributions coming from halo exclusion. We'll come back to these points in Section~\ref{sec:PkWithBiasRunning}.


The difference between the two implementations of sampling techniques (shuffling and mean) is that the one-point stochastic scatter about the mean is preserved in the shuffling method. When computing the cross power spectrum this stochastic noise does not contribute.  But the stochastic noise may contribute a small amount of white noise to the halo auto power spectrum. However, we argue in Appendix \ref{sec:ShufflingStochasticNoise} that the shuffling procedure destroys the contribution of this stochastic noise to the power spectrum.  This agrees with what we see by comparing the auto halo power spectrum from the shuffling and the mean methods.

Note that $b_{\rm h}$ in Fig.~\ref{fig:bh_Sim_Shuffle_Mean_800} is qualitatively similar to $b_{\rm c}$ in Fig.~\ref{fig:bc_Sim_Shuffle_Mean_800}, in particular as the smoothing scale increases,  $b_{\rm h}$ from the sampled field decreases and approaches that from the original simulation. We note that the $b_{\rm h} $ and $b_{\rm  c} $ from sampling agree with each other at large scales for large smoothing scales such as $50\, \mathrm{Mpc}/h$ (modulo the small differences pointed out above). This is an important point, and we will reiterate it after we re-interpret local biasing prescription.  Because of the problem of shot noise subtraction in the original (pre-shuffled) simulation, in this paper we shall mainly rely on the cross power spectrum when we compare results to simulations, except in Section \ref{sec:PkWithBiasRunning} where we address the low-$k$ regime of the auto power spectrum.

In order to shed light on the shuffling results, in Section \ref{sec:PkFromPT}, we shall apply perturbation theory to the local model. We shall see that the large-scale power problem for a halo biasing prescription with a small smoothing scale  already appears in the conventional test of local biasing prescriptions using perturbation theory. However, since the problem is particularly obvious for small smoothing scales, and one may worry about the accuracy of fitting the bias parameters in the scatter plot and the breakdown of perturbation theory when the variance is not small. The sampling technique introduced here evades these issues, and makes clear that this is a real problem of the standard formulation of the local biasing prescription.  We will show that the seemingly chaotic results from sampling in Fig.~\ref{fig:bc_Sim_Shuffle_Mean_800} and \ref{fig:bh_Sim_Shuffle_Mean_800} can be brought into nice agreement with the original simulation results when a new interpretation is offered in Section \ref{sec:PkWithBiasRunning}.


\section{ Power spectrum from perturbation theory: Local Model }
\label{sec:PkFromPT}

It is useful to compare the numerical power spectrum we obtained through sampling against those obtained from standard perturbation theory~\cite{PTreview}. We shall use the standard perturbative expansion with smoothing \cite{Heavens1998} because it is most directly related to our procedures. From the numerical results in Section \ref{sec:numerical_results}, it is already clear that they are qualitatively similar to the well-known results that when higher order loop corrections are included, they generally give rise to significant corrections even to large scales, unless the smoothing scales are sufficiently large \cite{Heavens1998}. There is an alternative procedure making use of the idea of renormalization to write these results in a different way \cite{McDonaldReBias}. However, after renormalization, the bias parameters become free parameters, and cannot be related to other bias parameters such as those in the scatter plot. Our ultimate goal is to establish consistent links between the bias parameters obtained from different means, such as the scatter plot and the power spectrum, therefore we shall not consider the specific scheme of renormalized bias advocated in~\cite{McDonaldReBias}. But we will comment on it more in Section \ref{sec:PkWithBiasRunning} since it has distinct predictions from our scheme.

In standard perturbation theory with smoothing, the nonlinear matter density field is smoothed. The perturbation series is expanded in terms of the linear matter power spectrum. When the next to tree-level correction is considered, all the 1-loop correction terms should be included. Here we summarize the results of the perturbative expansion of the halo cross power spectrum and auto power spectrum up to 1-loop, which will be needed later on.

\begin{figure*}[!t]
\centering
\includegraphics[ width=0.8\linewidth]{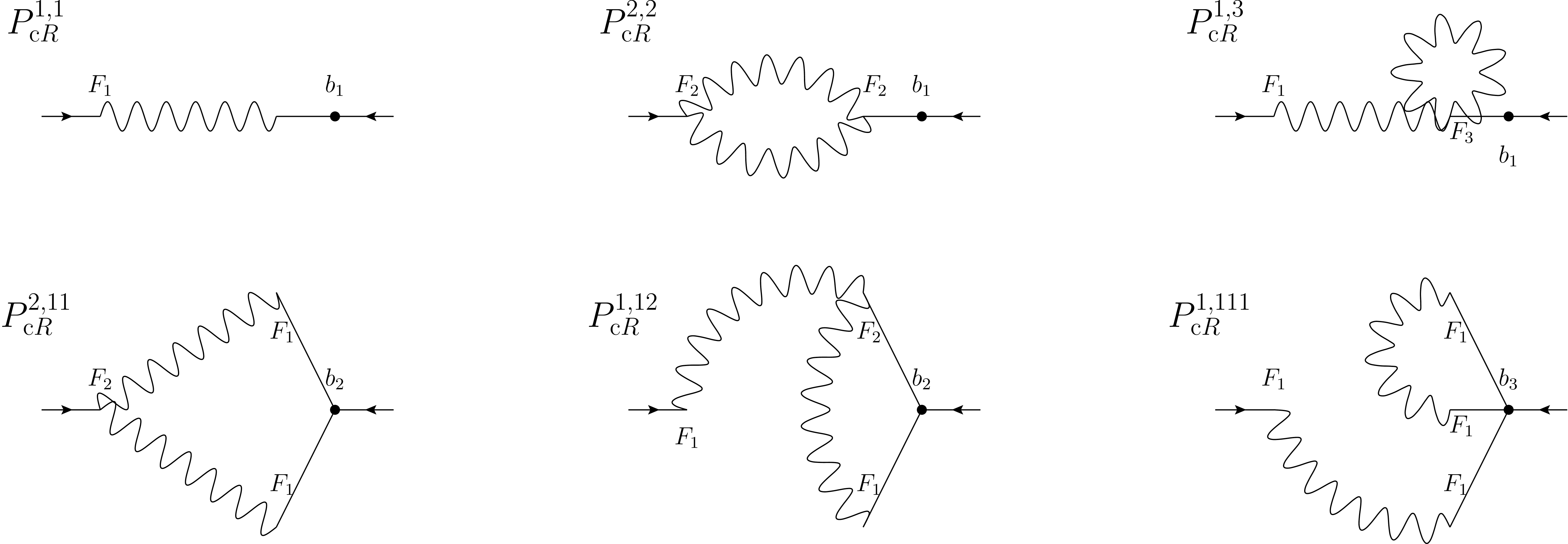}
\caption{ The diagrammatic representation of the perturbative expansion of the halo cross power spectrum up to 1-loop order.  See the text for the meaning of the notation. }
\label{fig:pc_feyndiag} 
\end{figure*}

\begin{figure*}[!t]
\centering
\includegraphics[ width=0.8\linewidth]{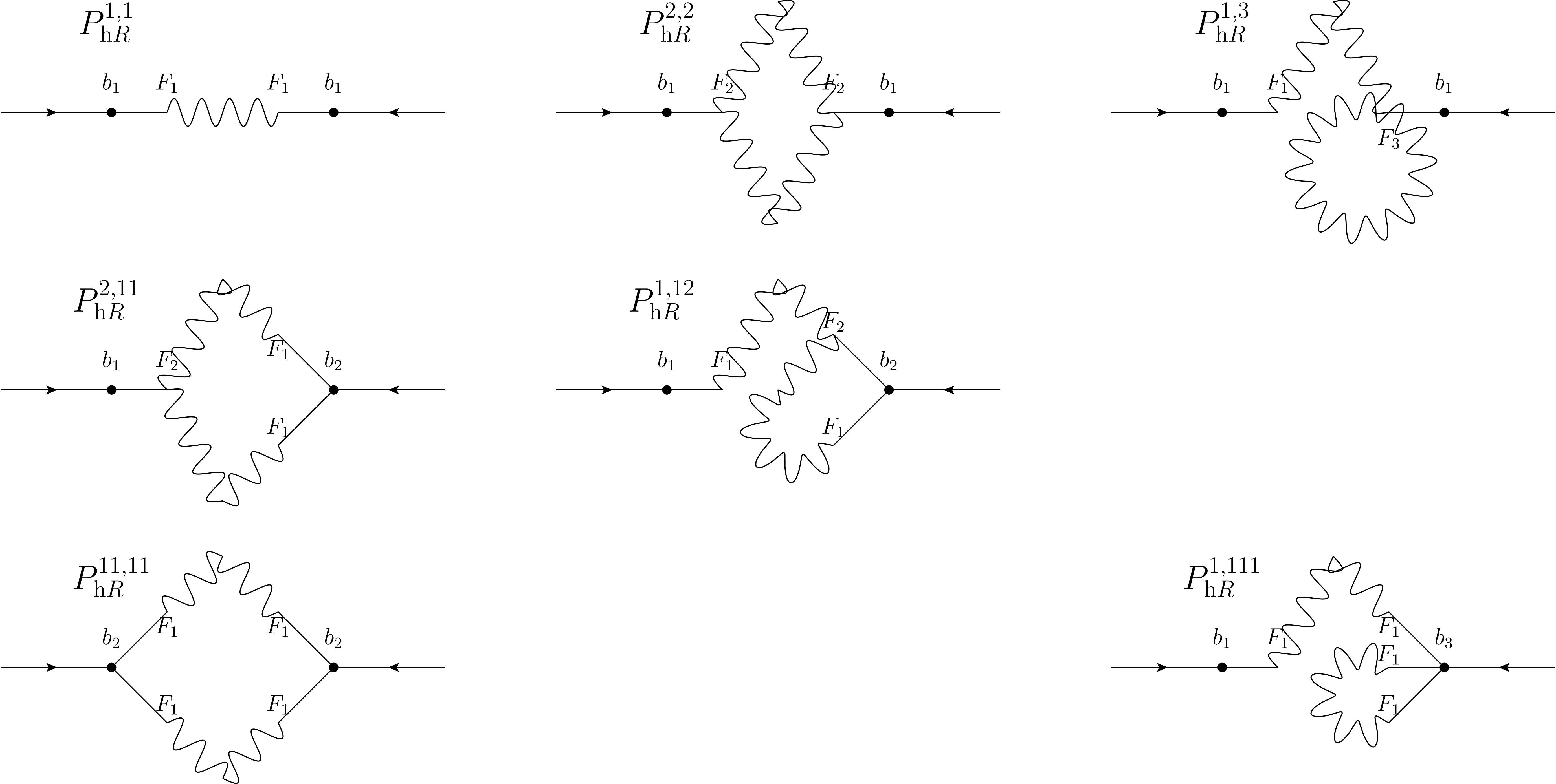}
\caption{ The diagrammatic representation of the perturbative expansion of the halo power spectrum up to 1-loop order.   }
\label{fig:pk_feyndiag} 
\end{figure*}

The perturbative expansion for the halo cross and auto halo power spectrum up to 1-loop order can be represented diagrammatically as in Fig.~\ref{fig:pc_feyndiag} and ~\ref{fig:pk_feyndiag} respectively. A diagrammatic representation for the biased density field has been used before, see \textit{e.g.}~\cite{Matsubara1995, Baldaufetal2011,MatsubaraNLBias}.  In each diagram, each line with arrow represents an external wave vector, a solid dot denotes a bias parameter vertex, and the branching of a solid line into wavy lines represents the vertex function due to the nonlinearity of gravity. Each wavy line represents the linear density field, and the merging of two wavy lines stands for a linear matter power spectrum. The meaning of our notation for the power spectrum can be illustrated by the example, $P_{\mathrm{c} R}^{1,12}$. The digit in front of the comma is for the dark matter field, while the number of digits behind the comma denotes the degree of biasing in the halo biasing prescription. In our example, it means it is of the form  $\langle \delta_R  b_2 \delta_R^2 \rangle $, in particular the degree of biasing is quadratic ($b_2$) as there are two digits behind the comma.  The numerical value of each digit denotes the order of each $\delta_R$ in the dark matter perturbation theory. In the example, the first factor (dark matter field) is linear, while the second factor (the halo part) contains a linear and a second order dark matter field. For the halo auto power spectrum, the notation is similar except that we have the dark matter field replaced by the halo field.

 The nonlinear effect of gravity is represented by wavy lines, we can imagine shrinking the wavy lines, or ``springs'', to a point, and so we see that the effect of gravity is to generate an effective 2-point function (the first rows in Fig.~\ref{fig:pc_feyndiag} and \ref{fig:pk_feyndiag}), effective connected 3-point function (the first two diagrams in the second row of Fig.~\ref{fig:pc_feyndiag} and the second row in Fig.~\ref{fig:pk_feyndiag}) for the nonlinear density field, and a connected 4-point function ($P_{\mathrm{h} R}^{11,11}$). The terms with $b_3^R$ gives rise to a product of power spectra. 

Here we shall consider quadratic and cubic local biasing models, \textit{i.e.}~$i$ runs up to 2 and 3 respectively in Eq.~(\ref{eq:EulerianLocalBiasing}). Using perturbation theory, or simply reading off from Fig.~\ref{fig:pc_feyndiag} with the appropriate symmetry factors,  we can expand the halo cross power spectrum as follows: 
\begin{eqnarray}
\label{eq:Pc1_1}
P_{\mathrm{c} R }^{b_1^R} (k )   & = &  b_1^R   W^2 (kR) P(k) = P_{\mathrm{c}  R}^{1,1} + P_{ \mathrm{c} R}^{2,2} +   P_{\mathrm{c} R}^{1,3}   , \\ 
\label{eq:Pc2_11}
P_{ \mathrm{c}  R}^{2,11}  ( k )   & = &    b_2^R   W(kR) \int \diff^3 q F_2(\mb{q}, \mb{k} - \mb{q} )  \\ 
&  \times &   W(qR)W( | \mb{k} - \mb{q} | R ) P_{\rm L}(q) P_{\rm L}(| \mb{k} - \mb{q} | ), \nonumber  \\
\label{eq:Pc1_12}
P_{ \mathrm{c} R}^{1,12}  ( k )   & = &  2   b_2^R   P_{\rm L}(k) W(kR) \int \diff^3 q F_2(-\mb{k}, \mb{q} )  \\ 
 & \times &  W(qR)W( | \mb{k} - \mb{q} | R ) P_{\rm L}(q),   \nonumber  \\ 
\label{eq:Pc1_3}
P_{\mathrm{c} R}^{1,111} ( k )   & = &   \frac{b_3^R}{2}  W^2(kR) P_{\rm L}(k)   \sigma^2_R   ,   
\end{eqnarray} 
where the mode coupling kernel $F_2$ is given by 
\begin{equation}
\label{eq:F2kernel}
F_2(\mb{p}, \mb{q} ) = \frac{5}{7} + \frac{1}{2} \mu  \big(  \frac{p}{q} +  \frac{q}{p} \big) + \frac{2}{7} \mu^2 ,
\end{equation} 
with $\mu = \hat{\mb{p}} \cdot \hat{\mb{q}} $, and $ \sigma^2_R $ denotes the linear RMS fluctuation 
\begin{equation}
\sigma^2_R =  \int \diff^3 q W^2(qR) P_{\rm L}(q). 
\end{equation}
Eq.~(\ref{eq:Pc1_1}) consists of the three terms proportional to $b_1^R$, corresponding to the three diagrams in the first row in Fig.~\ref{fig:pc_feyndiag}.  We have grouped these terms $ P_{\mathrm{c} R }^{1,1}$, $ P_{\mathrm{c} R}^{2,2}$ and $ P_{ \mathrm{c} R}^{1,3} $ together, as these three terms can be resumed to one term if we use the nonlinear matter power spectrum $P$, instead of the linear $P_{\rm L}  $. To do so, we shall use the nonlinear matter power spectrum $P$  measured from simulation. Thanks to the property of the $F_2$ kernel, Eq.~(\ref{eq:Pc2_11}) vanishes as $k$ goes to zero. 

If the window functions are not present, Eq.~(\ref{eq:Pc1_12}) can be computed analytically, and it is equal to $(34/21) b_2^R  \sigma^2  P_{\rm L} $, which is proportional to $P_{\rm L} $ and  is still significant at large scales. This must be approximately true in the presence of the window functions. Thus this term contributes to the excess power at large scales when the smoothing scales are not small.  Eq.~(\ref{eq:Pc1_3}) is also proportional to $P_{\rm L} $, thus this is another contribution to large scale power from loop corrections.  For the quadratic model, the term $ P_{\mathrm{c} R}^{1,111} $ vanishes by fiat.

We can carry out a similar expansion for the halo auto power spectrum: 
\begin{eqnarray}
\label{eq:Ph1_1}
P_{\mathrm{h} R }^{b_1^R}    (k )   & = & (b_1^R)^2  W^2 (kR) P(k) = P_{\mathrm{h}  R}^{1,1} + P_{ \mathrm{h} R}^{2,2} +   P_{\mathrm{h} R}^{1,3}   , \\ 
\label{eq:Ph2_11}
P_{ \mathrm{h}  R}^{2,11}  ( k )   & = &   2 b_1^R  b_2^R   W(kR) \int \diff^3 q F_2(\mb{q}, \mb{k} - \mb{q} )  \\ 
&  \times &   W(qR)W( | \mb{k} - \mb{q} | R ) P_{\rm L}(q) P_{\rm L}(| \mb{k} - \mb{q} | ), \nonumber  \\
\label{eq:Ph1_12}
P_{ \mathrm{h} R}^{1,12}  ( k )   & = &  4 b_1^R  b_2^R   P_{\rm L}(k) W(kR) \int \diff^3 q F_2(-\mb{k}, \mb{q} )  \\ 
 & \times &  W(qR)W( | \mb{k} - \mb{q} | R ) P_{\rm L}(q),  \nonumber  \\ 
\label{eq:Ph11_11}
P_{\mathrm{h}  R}^{11,11} (k )   & = &  \frac{ (b_2^R)^2 }{2} \int \diff ^3 q W^2(qR)  P_{\rm L}( q)   \\
 & \times &   W^2(| \mb{k} - \mb{q} | R) P_{\rm L}(| \mb{k} - \mb{q} |), \nonumber  \\ 
\label{eq:Ph1_3}
P_{\mathrm{h} R}^{1,111} ( k )   & = &  b_1^R  b_3^R  W^2(kR) P_{\rm L}(k)   \sigma^2_R . 
\end{eqnarray} 
The terms are similar to those for the cross power spectrum, except that there is a new term Eq.~(\ref{eq:Ph11_11}). For small $k$, one can expand the integrand in Eq.~(\ref{eq:Ph11_11}) about $\mb{k}$, and the leading correction to the constant term is $k^2$ only. Thus this term is rather constant at large scales, and its effect is like a shot noise term \cite{Heavens1998, McDonaldReBias}.   Again, we can resum the terms proportional to $ (b_1^R)^2$, $ P_{\mathrm{h} R }^{1,1}$, $ P_{\mathrm{h} R}^{2,2}$ and $ P_{ \mathrm{h} R}^{1,3} $, using the nonlinear matter power spectrum.

In order to consistently compare with the numerical results, we shall determine the bias parameters by fitting to the scatter plot between $\delta_R $ and $\delta_{\mathrm{h} R }$. There are two natural ways to fit to a scatter plot. One can fit simply to all the data points in the scatter plot. In this way, the fitting is roughly weighted by the number of data points, and so the fitting is dominated the region in the vicinity of the origin.   Alternatively, one can bin the points based on its value of $\delta_R$, and compute the mean and spread of $\delta_{\mathrm{h} R}$ in each bin. Fitting is made to the mean of the bins using the spread as the error bar.  In this way, we fit to the global shape of the $\delta_R-\delta_{\mathrm{h}R}$ distribution. \textit{A priori} it is not clear which way is the best approach. Sometimes, these two methods result in small but significant differences. Our goal here is to determine the halo power spectrum using the 1-point distribution determined from the scatter plot and the power spectum of the dark matter.

 When the smoothing scale is large, \textit{e.g.}~100 Mpc/$h$, the scatter in the scatter plot is small and the range of $\delta $ is narrow, and it is well fitted by a quadratic polynomial. In fact there is significant degeneracy between the bias  parameters. However, we find that the ones obtained by fitting to all the data points directly, \textit{i.e.}~simply weighted by the number of data points, recovers the bias parameter in the power spectrum most closely, which is simply equal to $b_1^R$ for such large smoothing scale. The one obtained by fitting to the global shape underestimates the bias parameter in the power spectrum sightly. Of course this does not guarantee that this is still the best approach in the weakly nonlinear regime. Nonetheless, we have checked that both methods give qualitatively similar results in recovering the bias parameters in the power spectrum in the weakly nonlinear regime. Therefore we adopt the method of fitting to all the data points in this paper.

\begin{figure*}[!t]
\centering
\includegraphics[ width=\linewidth]{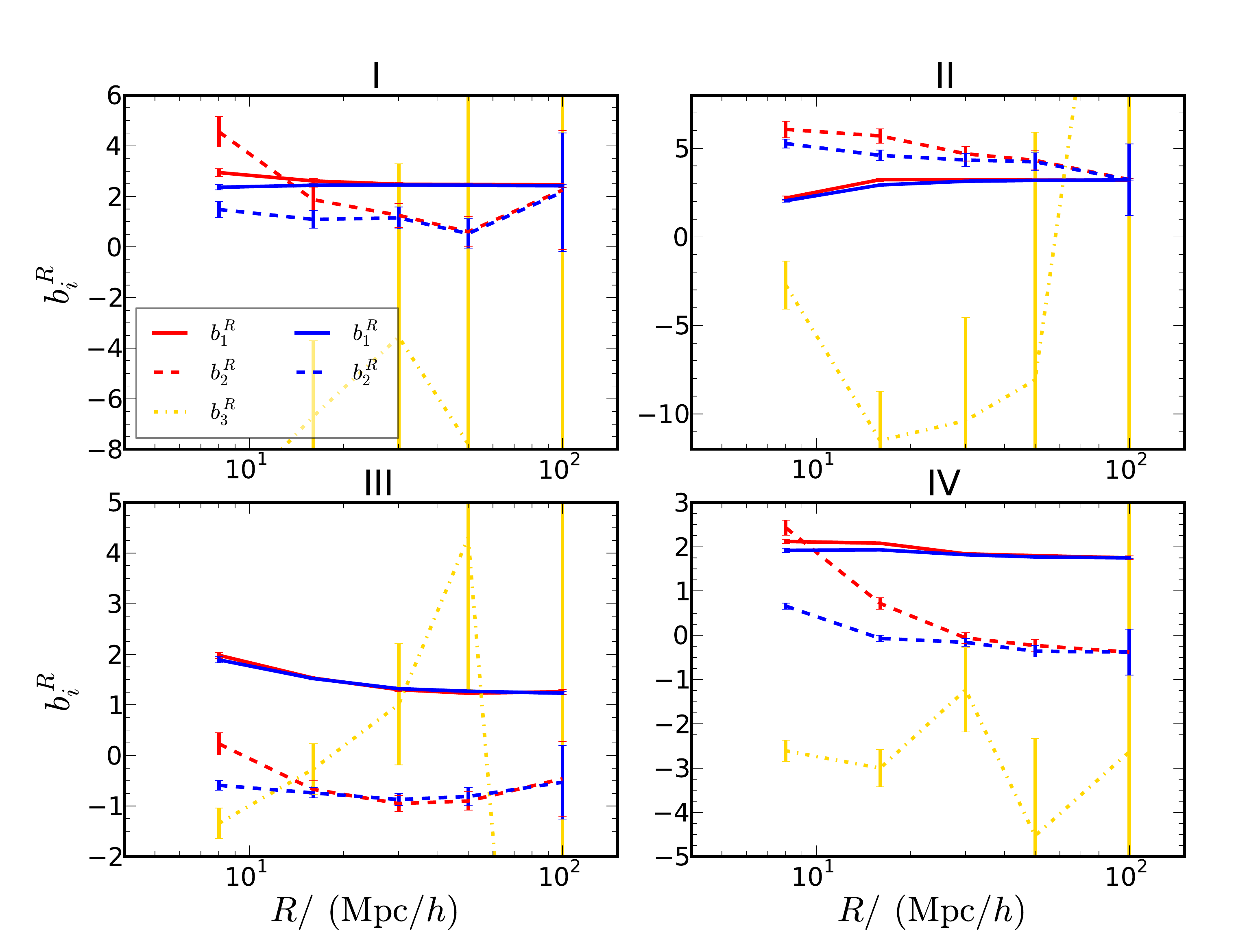}
\caption{ The bias parameters for the quadratic and cubic biasing models as a function of the smoothing scale.  The blue lines are for the quadratic model, while for the sake of clarity, we have used the red lines ($b_1^R$ and $b_2^R$) and yellow lines ($b_3^R $) for the cubic model. The error bars here are 2 $\sigma $.  The bias parameters generally run with the smoothing scale $R$. The lower the order of the bias parameter, the smaller the running. For $R\gtrsim 30 $ Mpc/$h$, the running of the bias parameters is generally small except perhaps for $b_3^R$ (which unfortunately becomes increasingly difficult to measure at large smoothing scales). }
\label{fig:running_bias} 
\end{figure*} 

To eliminate the influence of the outlier points, we have removed the data points whose $\delta_R$ falls within the largest or smallest 2\%.   In Fig.~\ref{fig:running_bias}, we show the best fit bias parameters for various halo groups using five different smoothing scales. We have considered both the quadratic and cubic biasing models.   We note that the bias parameters generally  ``\textit{run}''  as the smoothing scale varies. The higher the order of the bias parameter the more drastic the running it is. Similar results for the quadratic model have been reported in \cite{ManeraGaztanaga2011}. Detailed comparison is difficult as the halos are not exactly the same, but \cite{ManeraGaztanaga2011} reported that $b_2^R $ starts to approach 0 as $R$ decreases when $R \lesssim 20 \,  \mathrm{Mpc}/h$, which is at odds with our results. 

We note that the higher the degree of the bias parameter, the larger its error bar, as expected. 
In particular, the error bars of $b_3^R $ are so large in some cases for $R \gtrsim  50 $ Mpc/$h$ that it is consistent with $b_3^R$ being 0. We  expect that there is large degeneracy between the high degree bias parameters when $R$ is large because for large $R$, $\sigma_R^2$ is small, the data points are in close proximity of $\delta_R =0$ only. Thus $b_3^R $ cannot be reliably measured for large $R$.


In ~\cite{RothPorciani}, a Gaussian smoothing window instead of top-hat is used, and they generally do not use multiple window sizes to measure the same halo group, so direct comparison is not possible.  However, we note that using a Gaussian window, which has a weaker tail than top-hat window in Fourier space, results in milder running of bias parameters. Ref.~\cite{RothPorciani} argued that third degree biasing decription is much favoured than the quadratic one using the AIC information criterion. Here the large error for $b_3^R$ for large smoothing suggests that the data is not constraining on this additional parameter, and it may not be necessary. When the smoothing scale is small, the error bar is small, and so one may think that including $b_3^R$ is justified. However, we will see that the presence of $b_3^R$ often  causes larger discrepancies with the $N$-body results when they are used to compute the halo power spectrum.

\begin{figure*}[!t]
\centering
\includegraphics[ width=0.9\linewidth]{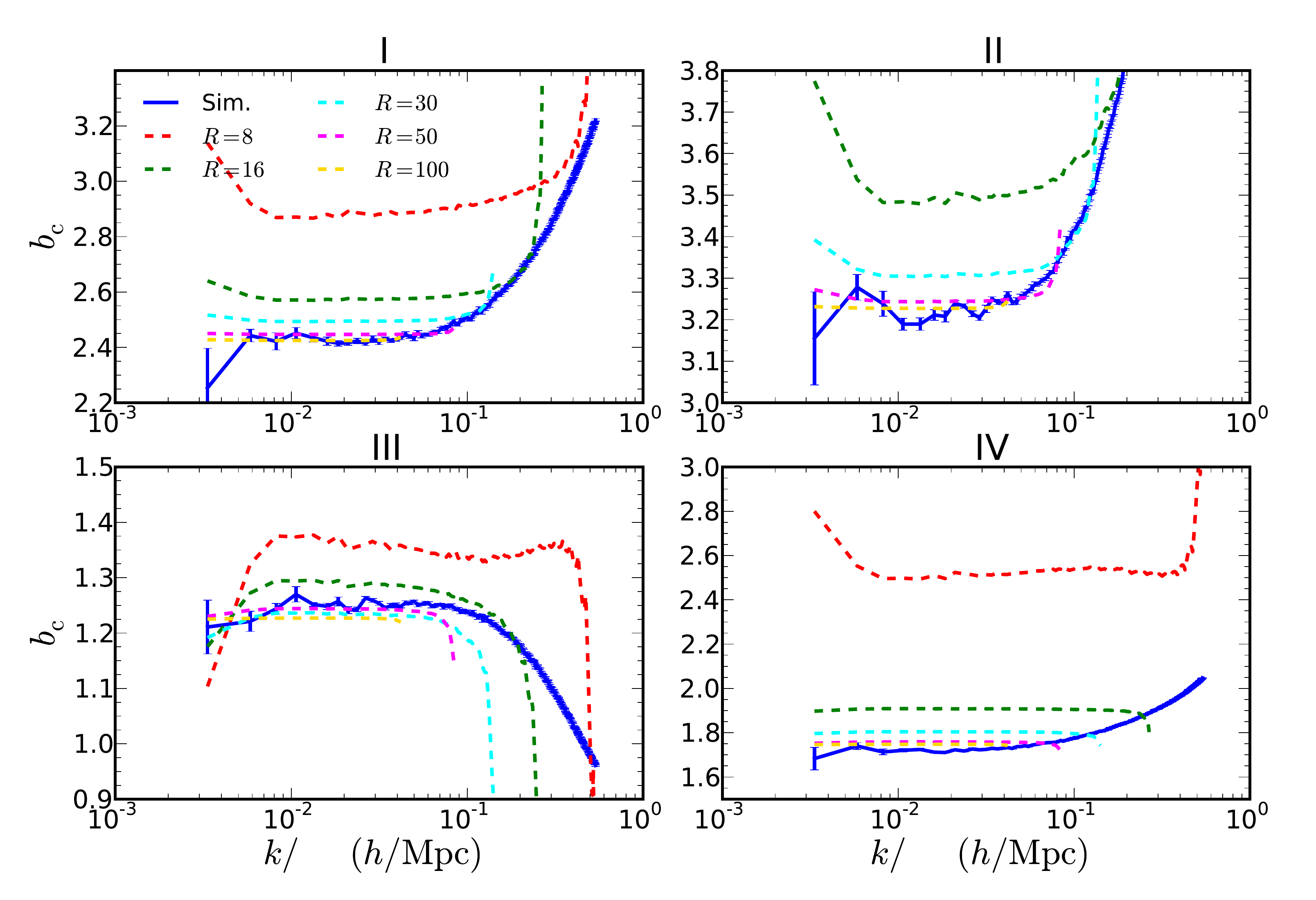}
\caption{ Comparison of $b_{\mathrm{c}}$ computed from the quadratic local biasing model compared to simulations (solid lines). The bias parameters are measured from the scatter plot and then used to make predictions, assuming a fixed smoothing scale $R$ as shown. The perturbative calculation includes all the one-loop corrections in the quadratic bias model. Each curve is shown up to the scale at which it starts to oscillate due to the window function.  Note the similarity of the ``failure" of these predictions with the sampling results in Fig.~1.  }
\label{fig:bc_PT2} 
\end{figure*}

With the bias parameters fitted from the scatter plot,  we can now compute the halo cross power spectrum and the auto power spectrum using  Eqs.~(\ref{eq:Pc1_1}--\ref{eq:Pc1_3}) and Eqs.~(\ref{eq:Ph1_1}--\ref{eq:Ph1_3}) respectively. In Fig.~\ref{fig:bc_PT2} we compare the $b_{\rm c}$ obtained from the original simulation and the one from perturbation theory using the bias parameters from a quadratic local biasing model. As $R$ increases, $b_{\rm c}$ from perturbation theory decreases at large scales and the agreement with the original simulation results improves. This is in qualitative agreement with that from sampling (Fig.~\ref{fig:bc_Sim_Shuffle_Mean_800}). When the cubic biasing model is used instead, the results are qualitatively similar to those from the quadratic model, except that for small smoothing scales, the deviations at large scales from the original simulation are slightly bigger.  Also for $b_{\mathrm{ h}} $, the results are also qualitatively similar to the those of  $b_{\mathrm{c}} $, and they are consistent with those from sampling method in Fig.~\ref{fig:bh_Sim_Shuffle_Mean_800}, so we do not show them here.

The perturbation theory and sampling ``fail'' to reproduce the original halo power spectrum in a similar way.  Using the sampling results, we can eliminate the two loopholes in the standard test of local biasing prescription, the inaccuracy of fitting to the scatter plot and the breakdown of perturbation theory. Thus we conclude that the problem originates from the standard interpretation of the local biasing prescription, where there is no relation of the smoothing scale $R$ to the ``observation scale" $k$. 

From the exercise in this section we have encountered two peculiar properties of local biasing related to the smoothing scale $R$: the scatter plot bias parameters run with $R$ and the power spectrum  varies with $R$ even at large scales. These two properties can be accommodated if we interpreted the smoothing scale $R$ as the ``knob'' that controls the scale of the fluctuations that the biasing prescriptions measure. We now discuss this in more detail.

\section{ Local biasing re-interpreted }
\label{sec:PkWithBiasRunning}
In Section \ref{sec:PkFromPT},  we have presented the standard procedure to compute the halo power spectrum using Eulerian local biasing with smoothing. This procedure  implicitly assumes that using the biasing relation obtained by smoothing the dark matter and halo fields at one scale $R$ is sufficient to compute the halo power spectrum at all scales. However, the window function of size $R$ filters out the small scale fluctuations, and so the local biasing relation gives the information mainly on the scale $R$.  This is reflected from the fact that for wave number $\gg R^{-1}$, the window function causes rapid decay of the power.

We have seen that in Section \ref{sec:PkFromPT} that there are two issues with the local biasing prescription related to the smoothing scale $R$. In  Fig.~\ref{fig:running_bias} we saw that the scatter plot bias parameters in fact are scale-dependent. We call this the \textit{running of bias}. In the standard approach outlined  previously, the running of bias is not incorporated. Rather, the scatter plot bias parameters are measured at one scale and applied to all scales. This assumes that 
\begin{equation}
\beta_i(R) = \frac{ \diff b_i^R }{ \diff \ln R   }
\end{equation}
vanishes. One may try to model the running of bias parameters, for example, using the peak background split or peak bias formalisms \cite{Desjacques08,DesjacquesSheth2010}.  A more practical way is to measure the scatter plot bias at various scales as in Fig.~\ref{fig:running_bias}.   Another would be to use a bispectrum analysis to get the running of bias parameters directly in Fourier space, but in this paper we are mostly interested in making connection with the literature where bias parameters are measured from scatter plots.

Another issue is that at large scales the power spectrum computed using the local biasing prescription depends on the smoothing scale \cite{Heavens1998}, as evident from Fig.~\ref{fig:bc_Sim_Shuffle_Mean_800}, \ref{fig:bh_Sim_Shuffle_Mean_800} and \ref{fig:bc_PT2}. This is a long-standing problem of using the local biasing prescription to compute the power spectrum. 

Here we argue that these two peculiar properties of the local biasing prescription call for a proper interpretation of the physical significance of the smoothing scale $R$. We propose that each biasing prescription specified by $R$ only enables us to calculate the Fourier space fluctuation at one scale corresponding to $R$. To our knowledge, this intuitive viewpoint on the physical significance of $R$ has not been appreciated (see however \cite{Scoccimarroetal2011}). The smoothing scale is generally regarded as unphysical, or one smoothing scale is sufficient. Perhaps, this is partly because theoretical predictions such as the peak-background split, at least as it is often implemented, give only one set of bias parameters. 

We interpret the local bias prescription obtained with smoothing window of scale $R$ as furnishing a good effective description \textit{at that scale.}  As any other effective description, it is valid only in certain regime, and we interpret the large discrepancy at large scales that we see in computing the power spectrum as a consequence that we apply the prescription outside its regime of validity. Taking this into account, we can naturally ``solve'' those two problems in the standard formulation of local bias, where there is no \textit{a priori} connection between the smoothing scale $R$ and the observation scale ($k$ in the power spectrum case).


Therefore, for each set of bias parameters measured from a scatter plot using a window of size $R$, we will use them to compute the power spectrum at a scale of roughly $1/R$ in Fourier space. Thus, each measurement of the bias parameters from a scatter plot only yields a data point in the power spectrum calculation, rather than the whole function as in the usual procedure. For the real-space top-hat window used in this work,  which is rather extended in Fourier space, we have some freedom in deciding the precise Fourier mode it corresponds to.  In this paper we regard the local biasing prescription as a large-scale effective description. On the other hand, the simulation results can be regarded as the ``full theory''. We can fix this degree of freedom by requiring that the effective description results match those of the ``full theory''.  We will see that the following mapping
\begin{equation}
\label{eq:kR}
k_R =  \frac{1.8} { R},
\end{equation} 
 results in remarkable agreement between the  sampling and the original simulations results.  For the power spectrum, the proper window to consider, is window function squared. It turns out that  the scale $k_R$ corresponds to the half-width-half-maximum of $W^2(kR)$.  We do not have a deeper justification for this prescription than what was just stated. However, there are cases in which this issue can be made more precise. 


Let us now discuss two examples where the mapping between the  smoothing scale $R$ and the observational scale ($k$ in the power spectrum) are obvious. The first would be to change the window function from a top-hat in real space, to a band power in Fourier space. That is, one can take a band-power window  given by
\begin{equation}
\label{eq:Wkshell}
W_{k, \Delta k}(q ) =  \big[  \theta(q - k +  \frac{ \Delta k }{2}  ) -   \theta(q - k -  \frac{ \Delta k }{2}  )  \big],
\end{equation}
which extracts a shell of width $\Delta k$ centering about $k$. Thus the scatter plot between $\delta$ and  $\delta_{\rm h}$ smoothed with the window $W_{k, \Delta k}(q )$ gives the biasing prescription describing the fluctuations of the halo density field at $k$. In this case, the mapping between the window smoothing scale and the observation scale in the power spectrum is completely fixed, as both are Fourier space quantities. However, the window Eq.~(\ref{eq:Wkshell}) is not convenient because it is highly nonlocal in real space and cannot be normalized in usual way, \textit{i.e.}~its integral over all the space vanishes.  The second example, is to keep the window a top-hat in real space, but calculate correlators in real space, in particular the variance of counts in cells at scale $R$. Then it's obvious as well that the mapping between smoothing scale $R$ and observation scale $R$ is one-to-one as well, there is no freedom at all.   In other words, note that our ``new" interpretation, when formulated in real space for count-in-cells statistics is nothing else that the standard well-known expression in Eq.~(\ref{eq:EulerianLocalBiasing}) for local bias originally put forward by Fry \& Gazta\~naga \cite{FryGaztanaga}.

In addition, note that loop corrections when calculating count-in-cells statistics will be suppressed at large $R$, \textit{e.g.}~linear bias renormalization vanishes for large cells since for the variance of the halo field,
\begin{equation}
\sigma_{\rm {h} R}^2 =  b_1^2\, \sigma^2_R + \Big({b_2^2\over 2}+b_1b_2 S_3\Big) \sigma^4_R + {b_2^2\over 4} S_4 \sigma^6_R,
\label{varCIC}
\end{equation}
where we have kept only terms up to quadratic bias, and $S_3$ and $S_4$ are the usual skewness and kurtosis parameters. Clearly, as $R\to \infty$ the loop corrections coming from nonlinear bias are highly suppressed compared to the linear term (that is, linear bias renormalization vanishes), and there is no extra noise contribution coming from nonlinear bias (in the language of~\cite{McDonaldReBias} ``shot-noise renormalization" vanishes). In particular, since there is no extra noise generated by nonlinear bias in the variance of the halo field $\sigma_h^2$, there should be no extra noise in the low-$k$ halo power spectrum. Our ``re-intrepretation" of local bias in Fourier space obeys these two properties as well.



\begin{figure*}[!t]
\centering
\includegraphics[ width=0.9\linewidth]{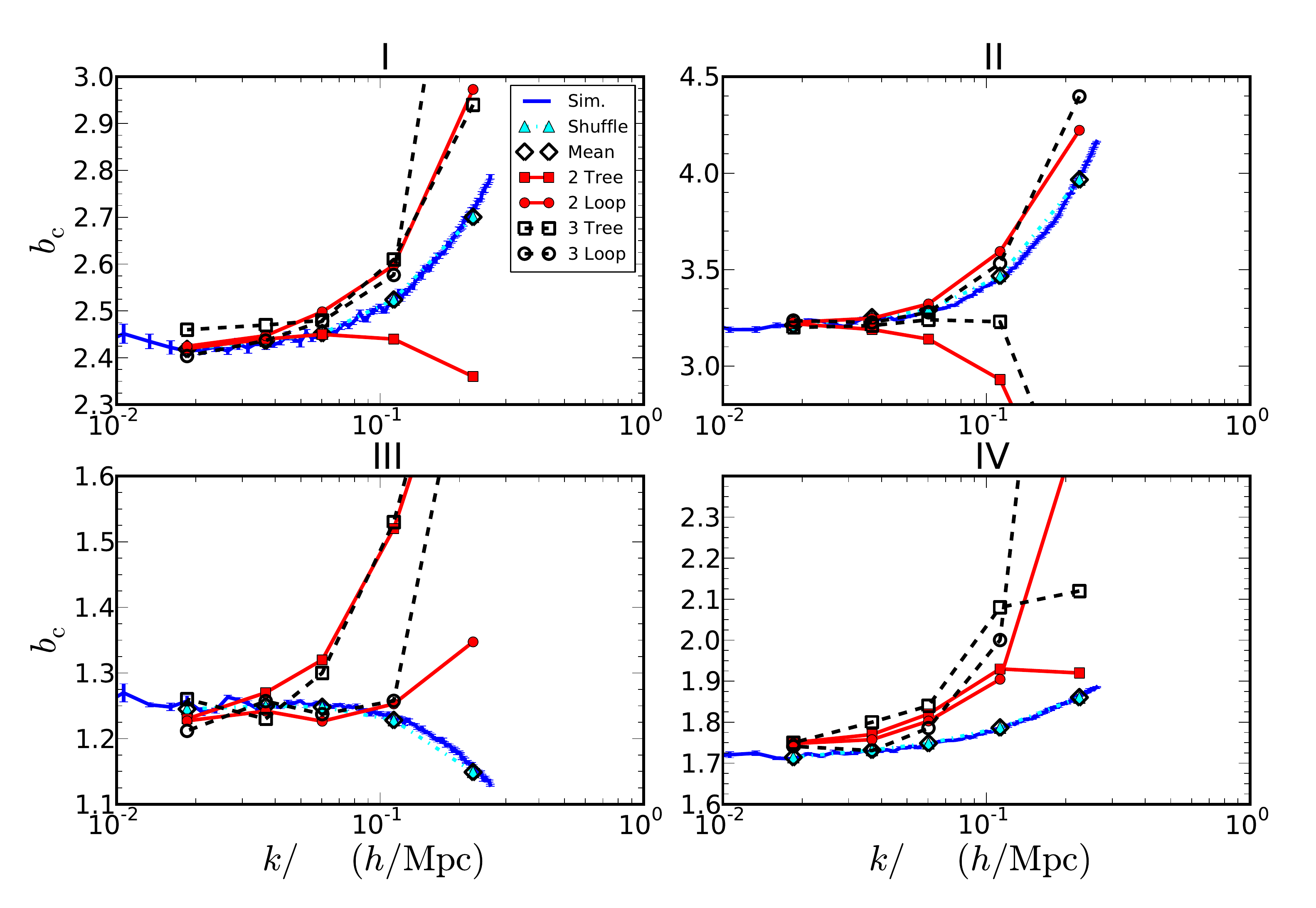}
\caption{  Comparison of the $b_{\rm c }  $ from the original simulation against the sampling and the perturbation  results for the quadratic and cubic model output only at $k_R $ given by Eq.~(\ref{eq:kR}) for each smoothing scale. The sampling results, both shuffling and mean, agree with the original simulation within 1\% (so it is difficult to differentiate them in the plot). For both the quadratic (solid symbols) and cubic (empty symbols) models, the 1-loop perturbation theory (circles) results generally improve the tree-level results (squares), except for the group IV.   Our interpretation of local bias solves the problems raised in Figs.~\ref{fig:bc_Sim_Shuffle_Mean_800} and~\ref{fig:bh_Sim_Shuffle_Mean_800}. }
\label{fig:bc_SimShuffleMean_TreeLoop2_3} 
\end{figure*}

Let us now go back to address the two issues at hand with the standard interpretation of local bias, and how our prescription solves these problems. Note that because the bias parameters are running, strictly speaking, we have to integrate over the bias parameters in computing the loop corrections too. However, the bias parameters vary less rapidly than the power spectrum, and the window functions $W( |\mb{k}_R - \mb{q} | R)$  in the integrand ensures that the dominant contribution to the integral is close to $k_R$. Thus it is a reasonable approximation to take the bias parameters outside the integral and assign the values at the scale $k_R$ to them.

For each smoothing scale, our interpretation is that we should only take the point at $k_R$ from Fig.~\ref{fig:bc_Sim_Shuffle_Mean_800} for the power spectrum calculated with bias parameters obtained from the scatter plot smoothed at $R$. Doing this we thus construct the plot in Fig.~\ref{fig:bc_SimShuffleMean_TreeLoop2_3}. Both the points from shuffling and mean almost coincide with the $b_{\rm c} $ from the original simulation, thus solving the large-scale excessive power issue in Figs.~\ref{fig:bc_Sim_Shuffle_Mean_800} (and \ref{fig:bh_Sim_Shuffle_Mean_800}). Up to 0.22 $h$/Mpc, for the four different halo groups, the $b_{\rm c}$ from sampling deviates from the one from original simulation by less than 1\%. This result is not very sensitive to the choice of the factor in Eq.~(\ref{eq:kR}). For example, if 2 is chosen instead of 1.8 in Eq.~(\ref{eq:kR}), the maximum deviation between the sampling results and the simulation is still within 2\%.

In Figure~\ref{fig:bc_SimShuffleMean_TreeLoop2_3}, we have also shown the points at $k_R$ taken from Fig.~\ref{fig:bc_PT2} (the quadratic model) and the cubic model,  which we did not show explicitly before. 
Let's first focus on the quadratic model. Generally by adding the loop corrections to the tree-level results, the agreement between theory and simulation indeed improves, except for the halo group IV at $k \sim 0.2 \, h/  \mathrm{Mpc}$. At redshift 0.97, with loop corrections, the difference between theory and simulation is within 4\% up to $ k=0.11$ $h$/Mpc, and within 10\%  up to $ k=0.22$ $h$/Mpc. The agreement deteriorates at redshift 0, and the agreement is within 7\% up to $ k \sim 0.11 \, h/  \mathrm{Mpc}$.  For the group IV, it is the only case in which the tree-level results fair better than the one including loop corrections. The pathological behavior of the perturbation results may be correlated with the fact that $b_2^R$ changes sign as $R$ decreases, from negative to positive, for the halo group IV.  Overall as redshift decreases, the nonlinearity is stronger, and so standard perturbation theory is expected to break down at smaller $k $. Note that similar levels of disagreement have been seen before for the perturbative calculation, \textit{e.g.}~see~\cite{Smith2007}. Here the deviations are slightly larger since our linear bias from the scatter plot does not agree as precisely with the low-$k$ matter-halo power spectrum bias.  Of course, one could improve significantly the performance of the perturbative calculation by {\em fitting} for the bias parameters that best match the simulation power spectrum measurements. That would have little to say though on the issues we are interested in this paper.

Now we compare the quadratic model with the cubic one. In general, they are more or less similar, except that the cubic one is less stable and often fairs worse than the quadratic model at large $k$. For large smoothing scales, the error bar on $b_3^R $ is large, so it is consistent with zero. From Fig.~\ref{fig:bc_SimShuffleMean_TreeLoop2_3}, because of large smoothing, its contribution is negligible, and so it is similar to the quadratic model. However, for small smoothing scales, such as 8 $\mathrm{Mpc}/h$, from Fig.~\ref{fig:running_bias}, we see that the error bar of $b_3$ is small and  it is clearly negative.  However, from Fig.~\ref{fig:bc_SimShuffleMean_TreeLoop2_3}, we see that the cubic model deviates further away from the simulation than the quadratic model. 
This is likely because for small smoothing scale the validity of the non-linear local bias expansion is very limited. 
Thus we prefer the quadratic model to the cubic one because of its simplicity and stability.

\begin{figure*}[!t]
\centering
\includegraphics[ width=0.9\linewidth]{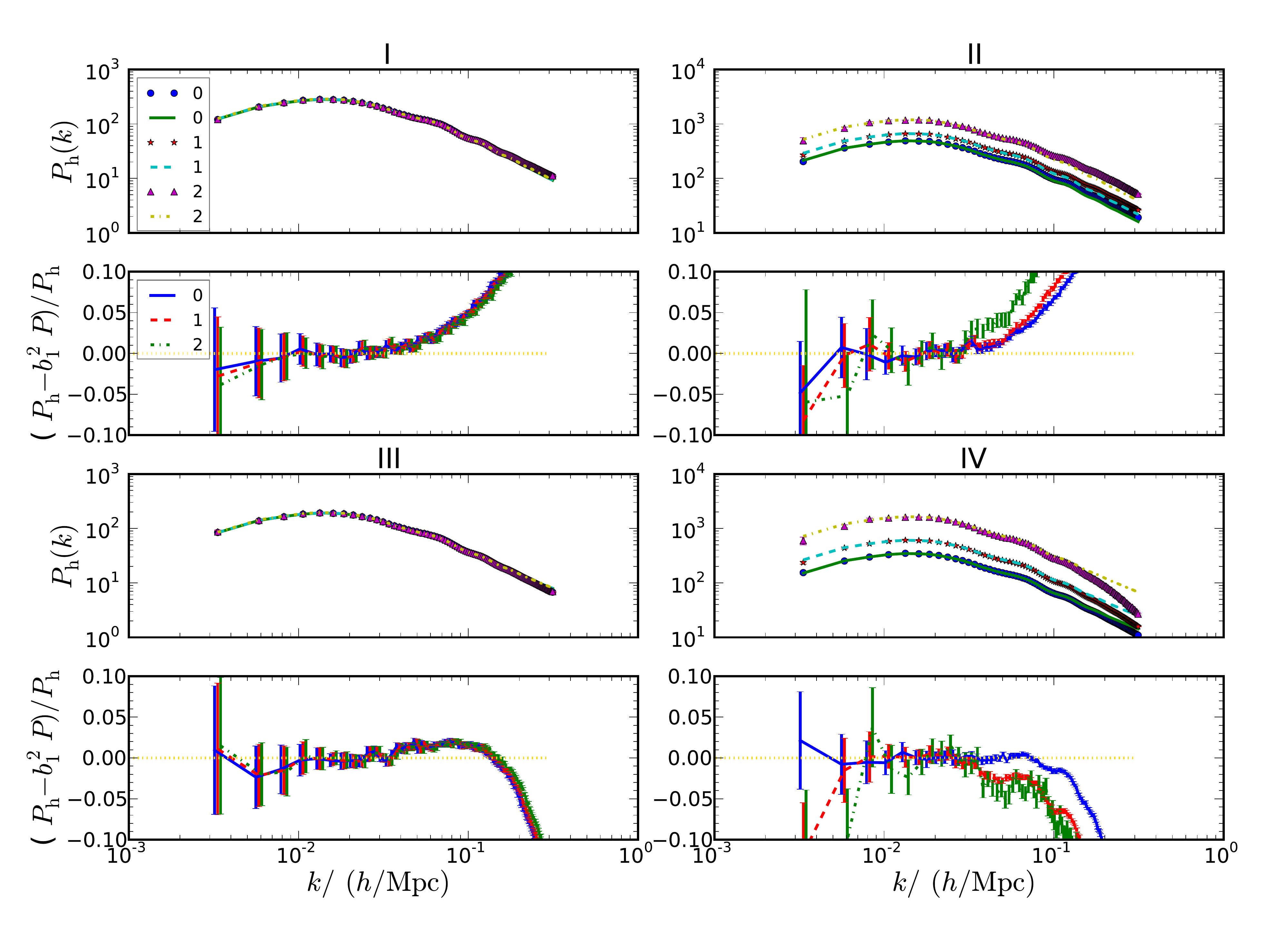}
\caption{The upper panels show the measured halo power spectrum from simulations (symbols) and the best fit linear bias model $b_1^2 P $ (lines) for each halo group. Three weighting schemes $m^k$, with $k=0$, 1 and 2 are shown. Mass weighting has little effects on the low mass groups, while it changes the power of the high mass groups.  The best fit $b_1$ is obtained by fitting to the simulation Poisson corrected halo power spectrum only up to $k_{\rm max } = 0.03 \, h/\mathrm{Mpc}$. The bottom panels show the residuals  in terms of fractional deviations $( P_{\rm h} - b_1^2 P ) /  P_{\rm h} $.  Up to $k \sim 0.004 \,  h/\mathrm{Mpc}$, for equal weighting, the simulation results are consistent with our prediction that the low-$k$ Poisson corrected halo power spectrum is well described by linear bias, with no evidence of shot-noise renormalization from nonlinear bias loops. For mass weighting by $m$ and $m^2$, we detect the sub-Poisson effects due to halo exclusion.   }
\label{fig:b1_LargeScaleFit_MW} 
\end{figure*}

Note that in our proposal there is a significant difference on the impact of loop corrections to the low-$k$ halo power spectrum. Indeed, our interpretation predicts that at large scales, the bias is simply given by $b_1^R$ with $R\to \infty$ and thus with negligible loop corrections. 
On the other hand, in the literature small smoothing scales are sometimes applied and the loop corrections are argued to be significant in the low-$k$ halo power spectrum, \textit{e.g.}~\cite{JimenezDurrer, Taruya08}. The reason for this, as discussed before, is that the term in Eq.~(\ref{eq:Ph11_11}) is quite close to constant at large scales, and so it resembles  shot noise \cite{Heavens1998,McDonaldReBias}. For non-Gaussian initial conditions of local type, similar loop corrections from nonlinear bias (proportional to $b_2^R  \sigma^2_R $) can lead to even stronger corrections at low-$k$ in the standard implementation of local bias~\cite{Taruya08}. However, as pointed out in \cite{Scoccimarroetal2011} such results rely  on using fixed smoothing scales  unrelated to the scale of fluctuations $k$.

At scales larger than the turnover of the halo power spectrum, any white noise starts to become increasingly important in the $k\to 0$ limit.  In \cite{JimenezDurrer}, it was argued that this shot noise effect maybe observable if the loop correction is large enough. Adopting our interpretation of local biasing, however, at $k \sim 0.005 \,   h/ \mathrm{Mpc}$ the corresponding smoothing scale should be of a few hundred Mpc, which makes the loop corrections  due to nonlinear bias vanishingly small. Even for 50 Mpc/$h$, the correction given by  Eq.~(\ref{eq:Ph11_11}) is of the order of $10^{-3}$--$10^{-2}$ of the tree-level term. 

Recall that in Section \ref{sec:numerical_results} we mentioned that $b_{\rm h} $ and $b_{\rm c} $ from sampling agree with each other at large scales when large smoothing scales are applied. With our interpretation of local biasing prescription, the sampling results indicate that $b_{\rm h}$ and $b_{\rm c}$ from local biasing prescription agree with each other.  In particular,  Eq.~(\ref{eq:Ph11_11}) has negligible effects at large scales.  We now provide additional tests of this property.


We performed an alternative test by fitting a linear bias model to the Poisson-subtracted halo auto power spectrum at large scales. Besides the standard equal weighting of all halos in a given mass bin, which is the only scheme that we used so far, we now also consider mass weighting with weight $m$ and $m^2$, where $m$ is the mass of the halo. Mass weighting enables us to preferentially weight the more massive and hence more biased halos in the sample. When the halos are weighted, the Poisson shot noise contribution is modified to Eq.~(\ref{eq:massweight}).  Mass weighting has little effect on the low mass halo groups since our mass bin is quite narrow, while it increases the auto power spectrum and the shot noise significantly for high-mass halo groups. In Fig.~\ref{fig:Ph_Pshot}, we show the halo auto power spectrum and the shot noise for each of the halo group.  In particular for mass weighting by $m^2$ the shot noise is higher than the signal at most of scales of interest ($k\lesssim 0.1 \,   h / \mathrm{Mpc}$). Also, mass weighting gives more weight to massive halos, which is expected to lead to larger halo exclusion effects. On the other hand, as the more biased objects are preferentially weighted, the magnitude of $b_2$ is expected to increase. The contribution due to Eq.~(\ref{eq:Ph11_11}) will increase too. However, according to our proposal, this term is still negligible at large scales even with mass weighting. 


 The results are presented in Figure~\ref{fig:b1_LargeScaleFit_MW}. The top panel shows the halo power spectrum with Poisson shot noise subtracted (data points), and a linear bias model  $  b_1^2~ P $ (solid lines), where $P$ is the measured matter power spectrum. The fit to $b_1$ is done using data up to $k_{\rm max } = 0.03 \, h/\mathrm{Mpc}$, safely within the linear bias regime for all halo samples. 



Because we subtracted Poisson shot noise, any shot noise renormalization at low-$k$ will show up as residuals from our fit. The bottom panels in Fig.~\ref{fig:b1_LargeScaleFit_MW} show the residuals in terms of the fractional deviations  given by $( P_{\rm h} - b_1^2 P )/  P_{\rm h} $. Up to $k \sim 0.004 \,  h/\mathrm{Mpc}$, for equal weighting ($m^0$ weighting) the simulation results are consistent with our statement that local biasing predicts that the $k \to 0$ power spectrum is given by the linear bias alone, \textit{i.e.}~there is no evidence for shot noise renormalization.   These test results alone only constrain that the sum of the correction by  Eq.~(\ref{eq:Ph11_11}), the sub-Poisson effects due to halo exclusion and any other stochastic noise are small.  The drop in power at large scales for mass weighting is consistent with the expectation that mass weighting increases the sub-Poisson effect due to halo exclusion.  We do not see any evidence for an {\em increase} in the noise level expected from nonlinear bias in the ``standard" interpretation of local bias as massive halos are up-weighted. If that contribution is there, it must always be somehow overcompensated by the negative noise coming from exclusion.

An alternative look for extra noise at low-$k$ can be obtained, as discussed in Section~\ref{sec:numerical_results}, by comparing cross and auto bias, $b_c$ and $b_h$. In our interpretation of local bias, these two should agree at low-$k$ modulo the effects of halo exclusion (which makes $b_h<b_c$), whereas in the ``standard interpretation" shot noise renormalization due to nonlinear bias would make $b_h>b_c$ (again, modulo halo-exclusion effects). Correcting for halo-exclusion effects is nontrivial, the two-point function of discrete halos with exclusion $\xi_d$ can be written as (see Appendix A2 in \cite{Smith2007})
\begin{equation}
1+\xi_d(r) = [1+\xi_h(r)][1+h(r)]+ {1\over \bar{n}}\delta_{\rm D}({\bf r})
\label{excl}
\end{equation}
where $\xi_h$ is the continuous version and $h$ is the  two-point function of random hard-spheres. In the approximation we are working in cosmology, where the number density $\bar{n}$ of halos is small (otherwise see \cite{HansenMcDonald, Wertheim1963,  Thiele1963}), we can approximate $h(r)=-1$ for $r<r_e$ and zero otherwise, where the exclusion radius $r_e$ is \textit{e.g.}~twice the halo radius when dealing with equal mass objects. This means that $h(r)=-W_{\rm TH}(r,r_e)$ where $W_{\rm TH}$ is the spherical top-hat window function and we have for the continuous power spectrum~\cite{Smith2007},
\begin{eqnarray}
P_h(k) &=& P_d(k)- {1\over (2\pi)^3}{1\over \bar{n}} +W_{\rm TH}(kr_e)\nonumber \\ 
& & + \int W_{\rm TH}(qr_e) P_h(|{\bf k}-{\bf q}|)d^3q
\label{exclF}
\end{eqnarray}
While we see straight away that short-range exclusion from hard-sphere halos contributes opposite to the Poisson noise from discreteness, there is an issue in that the correction for exclusion depends on the continuous power spectrum itself through the convolution term {\em at scales comparable and below the exclusion radius}. This term is the largest contribution due to exclusion (with the subdominant term $W_{\rm TH}(kr_e)$ becoming just the exclusion volume in the low-$k$ limit). In~\cite{Smith2007} $P_h$ in the convolution term is modeled by assuming the two-point function is a power-law at small scales, but this is not enough for our  purposes as any deviations from it will result in unaccounted noise, precisely what we want to test. It is  difficult to model $P_h$ close to the exclusion radius from first principles, e.g. local bias by definition breaks down at these scales. Therefore, we won't attempt to correct for exclusion effects in this paper. 

We can still look at low mass halos where exclusion effects are negligible and see whether we detect any extra noise, i.e. whether we see any evidence of $b_h>b_c$. The simulations we have presented so far have too poor resolution for this (the particle mass is $4.57\times 10^{11} M_\odot/h$) but we have also looked at higher-resolution simulations of the {\em same} cosmology (the Carmen runs in the LasDamas simulations) with particle mass $4.94\times 10^{10} M_\odot/h$. In these simulations, we have studied halos in different mass bins, as low as $10^{12} M_\odot/h$, with linear bias less than unity. We have found that the ratio $b_h/b_c$ is always within one percent of unity at the largest scales we can probe, of order $k=0.006 h/{\rm Mpc}$. For reference, we quote this ratio for the halo samples used throughout this paper: $b_h/b_c=1,0.98,1.015,0.99$ for halo samples I through IV.  When splitting the same halos in different bins, e.g.  $10^{13}-10^{14} M_\odot/h$ and above $10^{14} M_\odot/h$ we find $b_h/b_c=1,0.92$ at $z=0$ and $b_h/b_c=1,0.93$ at $z=1$. Thus while there is evidence for halo exclusion for high-mass halos (which also shows up when mass weighting the wider bins), only for sample III there is very slight tendency for $b_h/b_c$ to be above unity. However, when we look at the higher-resolution simulations of the {\em same} halos in sample III (that have between 162 and 304 particles rather than 20 to 36 particles in the low resolution runs), even this slight hint of a discrepancy is reduced to consistent with unity. This is perhaps not surprising, halos with as few as 20 to 40 particles in the low resolution run have bias factors that differ systematically from the same halos in the high resolution run by almost 10\%,  thus one should not trust their clustering properties at the percent level despite of their small statistical error bars.

\section{ A case study: power spectrum of a nonlocal model      }

\label{sec:NonlocalModelStudy}

The halo sampling method can be applied to detect the existence of extra nonlocal contributions to halo bias since the sampled halo field should result in the same halo density field if local bias holds.  On the other hand, some insights can be gained by studying a specific model. In this section, we compute the corrections to the power spectrum due to the nonlocal bias induced by gravitational evolution derived in \cite{NLBias} (see also \cite{Fry96, Catelan1998,McDonaldRoy,MatsubaraNLBias,Baldaufetal2012}). We will first give the expressions for the nonlocal corrections to the power spectrum. We then use the estimated values of the nonlocal bias parameters to illustrate the importance of these corrections to the halo power spectrum.  Since the local model is often assumed, we then go on to study the effects in the bias parameters and the resultant power spectrum when the underlying model is nonlocal but a local model is assumed.

\subsection{ Nonlocal bias contribution to the power spectrum }
\label{sec:NLPk}
In this section, we study the corrections to the power spectrum due to the nonlocal bias induced by gravitational evolution. 

We take the nonlocal bias model derived  in \cite{NLBias}. Let us first review some of the results that we will need here. We consider the simple model in which the initial bias is local and described by the  bias parameters $b_1^*$, $b_2^*$ and $b_3^*$ at time $t_*$, subsequent gravitational evolution will then induce nonlocal bias. At time $t>t_* $, the terms in the biasing prescription can be classified into two types: those present initially, \textit{i.e.}~the local type, and the nonlocal terms that are induced by subsequent evolution. Here we will focus on the nonlocal terms.  For simplicity we assume that halos are  conserved. When the halo conservation assumption is relaxed, and some local source function describing galaxy/halo formation and merging is introduced, the structure of the induced kernels remains the same as that derived from the conserved model (see~\cite{NLBias} for details).

At $t  \ge  t_*$, to the third order, the halo density is given by \cite{NLBias}
\begin{eqnarray}
\label{eq:NLbiasing}
\delta_{ \mathrm{h} }& =& b_1 \delta + \frac{b_2}{2} \delta^2  +   \frac{b_3}{6} \delta^3  + \gamma_2   \mathcal{G}_2( \Phi_{\rm v} )( 1 + \beta \delta )   \\  
& + &  \gamma_3  \big[ \mathcal{G}_3 ( \Phi_{\rm v} )  +  \frac{6}{7} \mathcal{G}_2 (\Phi^{(1)}_{\rm v} , \Phi_{\rm 2LPT}  )  \big] ,  \nonumber  
\end{eqnarray}
where $\mathcal{G}_2$ and $\mathcal{G}_3 $ are given by 
\begin{eqnarray}
\mathcal{G}_2 (\Phi_{\rm v}  )   & = &  ( \nabla_{ij} \Phi_{\rm v} )^2 - (\nabla^2 \Phi_{\rm v} )^2,    \\
\mathcal{G}_3 (\Phi_{\rm v}  )   & = & (\nabla^2 \Phi_{\rm v})^3 +   2 \nabla_{ij} \Phi_{\rm v}  \nabla_{jk} \Phi_{\rm v}  \nabla_{ki} \Phi_{\rm v}      \\
                    &- &   3 ( \nabla_{ij} \Phi_{\rm v} )^2 \nabla^2 \Phi_{\rm v}  ,      \nonumber  
\end{eqnarray}
with $\Phi_{\rm v} $ being the velocity potential.  $ \mathcal{G}_2 (\Phi_{\rm v} , \Phi_{\rm 2LPT }  ) $ is a variant of $\mathcal{G}_2 (\Phi_{\rm v} ) $ and it is defined as 
\begin{equation}
\mathcal{G}_2 (\Phi_{\rm v} , \Phi_{\rm 2LPT }  ) = \nabla_{ij} \Phi_{\rm v}  \nabla_{ij} \Phi_{\rm 2LPT} - \nabla^2  \Phi_{\rm v}  \nabla^2 \Phi_{\rm 2LPT},
\end{equation} 
with $\Phi_{\rm 2LPT}$ being the potential arising in second order Lagrangian perturbation theory given by 
\begin{equation}
\nabla^2   \Phi_{\rm  2LPT } = -  \mathcal{G}_2 ( \Phi_{\rm v} ) .
\end{equation}
 The nonlocal bias parameters $\gamma_2 $ and $\gamma_3$ are 
\begin{eqnarray}
\label{eq:gamma2}
\gamma_2 & = &   -\frac{2}{7}( b_1 -1 ) ( 1 - \frac{1}{g} ) ,\quad  \beta = \frac{b_2 }{b_1 - 1} , \\
\label{eq:gamma3}
\gamma_3 & = & \frac{ (b_1 -1) }{ 63 } \Big( 11 - \frac{18}{g} + \frac{7}{ g^2 } \Big),   
\end{eqnarray}
where $ g $ denotes the ratio of growth factor between the observation time and the formation time, $D/D_*$. We note that the values of these nonlocal bias parameters are model-dependent, but the tensorial structure of the nonlocal terms should be generic.  It is worth commenting on smoothing when these nonlocal terms are present.   Here we regard these nonlocal terms as extra degrees of freedom, in addition to density $\delta$. Hence the smoothing operator applies to $\mathcal{G}_2$ and  $\mathcal{G}_3$ in the same way as to $\delta$.

Using the prescription in Eq.~(\ref{eq:NLbiasing}), we can compute the correction to the power spectrum due to the nonlocal bias terms. Up to 1-loop, they are given by 
\begin{eqnarray}
\label{eq:Pb1gamma2_1}
 P_{b_1  \gamma_2 }^1(k) & = & 8  b_1 \gamma_2  P_{\rm L}(k) W^2(kR)  \int \diff^3 q    \\
& \times & ( \mu_{  \mb{k} -  \mb{q}, \mb{q}  }^2 - 1 )    G_2 ( - \mb{k} , \mb{q})  P_{\rm L}(q)         \nonumber   \\    
\label{eq:Pb1gamma2_2}
P_{b_1 \gamma_2}^2 (k)&=&  4  b_1 \gamma_2  W^2(kR) \int \diff^3 q F_2( \mb{q}, \mb{k} - \mb{q} )                   \\
& \times & ( \mu_{\mb{q}, \mb{k}-\mb{q}}^2 -1 )  P_{\rm L}(q) P_{\rm L}(|\mb{k} - \mb{q} |) ,      \nonumber             \\
\label{eq:Pb2gamma2}
P_{b_2 \gamma_2}(k)  &= & 2 b_2 \gamma_2 W(kR) \int \diff^3 q ( \mu_{\mb{q}, \mb{k}- \mb{q} }^2 -1 )                \\
& \times & W(qR) W(|\mb{k}-\mb{q}   | R) P_{\rm L}(q) P_{\rm L}(| \mb{k}-\mb{q} |),                 \nonumber             \\
\label{eq:Pgamma2gamma2}
P_{\gamma_2 \gamma_2 }(k) & = &  2 \gamma_2^2   W^2(kR)  \int \diff^3 q  ( \mu_{\mb{q}, \mb{k}- \mb{q}}^2 -1 )^2  \\
          & \times  & P_{\rm L}(q) P_{\rm L}(| \mb{k} - \mb{q} | )   ,                             \nonumber              \\   
\label{eq:Pb1gamma3_2LPT}
P_{b_1 \gamma_3}^{\rm 2LPT } (k)  & =&  - \frac{24  b_1 \gamma_3}{ 7} W^2(kR)  P_{\rm L}(k) \int \diff^3 q                        \\
  &\times &  ( \mu_{\mb{q}, \mb{k}+ \mb{q} }^2  -1 ) ( \mu_{\mb{k}, \mb{q} }^2 -1 ) P_{\rm L}(q) ,\nonumber       \\
\label{eq:Pb1gamma3}
P_{b_1 \gamma_3} (k)  &=& 0,               
\end{eqnarray}
where $ \mu_{\mb{q}, \mb{p}}  $ denotes $\mb{q} \cdot \mb{p} / (q p ) $. The meaning of each term should be self-evident. There are two terms proportional  to $b_1  \gamma_2 $.  Eq.~(\ref{eq:Pb1gamma3_2LPT}) is due to the $b_1$ term  and the $\gamma_3$ term with the 2LPT potential. There is another term proportional to $ b_1 \gamma_2 \beta $, but it is equal to $ (b_1 -1 )/ ( 2 b_2 ) P_{ b_2  \gamma_2 } $, so we do not show it explicitly here. In all the terms except for Eq.~(\ref{eq:Pb2gamma2}), the window function simply factors out. Interestingly, Eq.~(\ref{eq:Pb1gamma2_1}) and (\ref{eq:Pb1gamma3_2LPT}) can be further simplified to the same functional form up to a proportionality factor as 
\begin{eqnarray}
&      &   A  \frac{  P(k)  }{ k^3 }  W^2(kR)  \int \diff q \frac{P(q)}{q^3 }  \Bigg[ 4 k q (k^2 + q^2 )    \nonumber    \\
& \times  & ( 3 k^4 - 14 k^2 q^2 + 3 q^4 )            
+       3 (k^2 -q^2)^4 \ln \frac{(k-q)^2 }{(k+q)^2 } \Bigg]  ,   \nonumber     
\end{eqnarray}
with $A= ( \pi / 14 ) b_1 \gamma_2  $ for  $P_{b_1  \gamma_2 }^1 $ and  $A= ( \pi / 14 ) b_1 \gamma_3 $ for  $  P_{b_1 \gamma_3}^{\rm 2LPT }$.  From \cite{NLBias} ( and also Eq.~(\ref{eq:gamma2}) and (\ref{eq:gamma3}) ),  we expect $\gamma_2 $ to be generically negative, while $ \gamma_3$ is positive and smaller in magnitude, and so these two terms partly cancel each other.

\begin{figure}[!t]
\centering
\includegraphics[ width=0.9\linewidth]{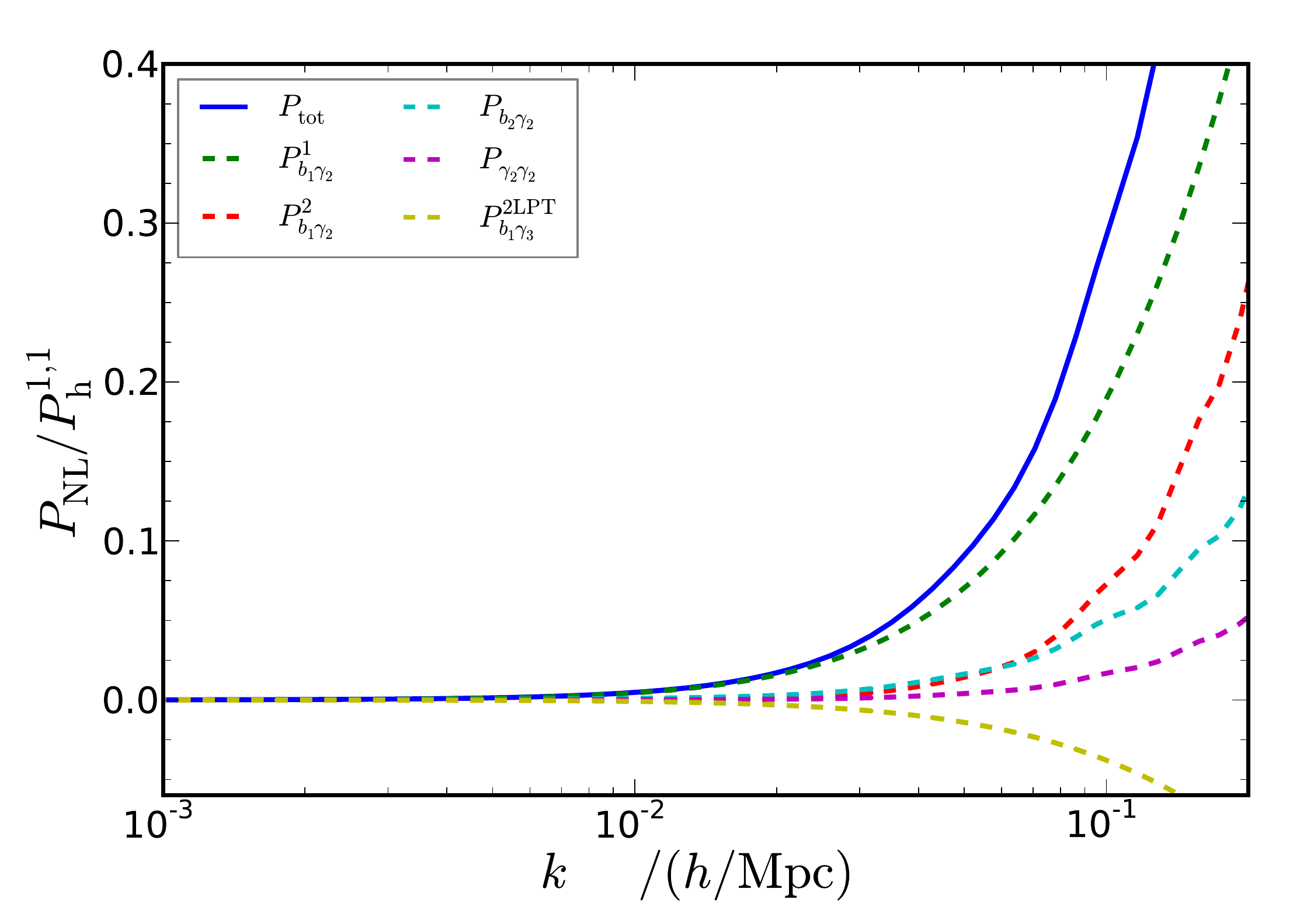}
\caption{ 
The total nonlocal bias contribution to the power spectrum (solid line) and its individual nonlocal contributions (dashed lines). The corrections to the halo power spectrum due to nonlocal bias, are normalized with respect to the linear bias term using a linear power spectrum $P_{\rm h }^{1,1} = b_1^2 P_{\rm L}   $. The bias parameters used are $b_1=2$, $b_2=1.34$, $\gamma_2= -0.5$ and $\gamma_3 = 0.1$.  The nonlocal corrections vanish at large scales and start to contribute significantly  at $k \sim 0.1 \, h/\mathrm{Mpc} $.   }
\label{fig:Pk_NLcomponent} 
\end{figure}

In Fig.~\ref{fig:Pk_NLcomponent}, we show the correction terms
Eq.~(\ref{eq:Pb1gamma2_1}--\ref{eq:Pb1gamma3_2LPT}) using the bias parameters
$b_1 = 2$, $b_2 = 1.34$, $\gamma_2=-0.5$, and $ \gamma_3 = 0.1  $. We have normalized the correction terms with respect to the tree-level term  $P_{\rm h}^{1,1} = b_1^2 P_{\rm L}   $, where $P_{\rm L}$ is the linear power spectrum.  At large scales, the contribution due to the nonlocal terms vanish. This is because the nonlocal kernels are derived from momentum conservation equations, and so they vanish when the external wave number goes to zero \cite{Goroff}.  At $k \sim 0.1 $ $h$/Mpc, the correction due to these nonlocal terms is $30$ per cent compared to $P_{ \rm h }^{1,1}$. The precise contribution depends on the nonlocal bias parameters, which should be obtained by fitting to the full model. Note that in making this plot we have used a window size of 16 Mpc/$h$, but the window function has negligible effect in the ratios shown in the figure.

Figure~\ref{NlocTOloc} shows the ratio of the effective local model (eq.~(\ref{eq:LFitModel}) below) power spectrum to the underlying nonlocal model. For the underlying nonlocal model, the local terms Eq.~(\ref{eq:Ph1_1}--\ref{eq:Ph11_11}) and the nonlocal terms  Eq.~(\ref{eq:Pb1gamma2_1}--\ref{eq:Pb1gamma3}) are used, and the bias parameters are the same as those in Fig.~\ref{fig:Pk_NLcomponent}. The effective local model includes the terms Eq.~(\ref{eq:Ph1_1}--\ref{eq:Ph11_11}) and the corresponding effective local bias parameters are obtained using Eq.~(\ref{eq:AveLBias}).  As local biasing prescription is a sensible approximation only when the scale dependence is properly taken into account, we have used our mapping of scales, Eq.~(\ref{eq:kR}). That is, each $k$ mode is computed using a different window function.  We see from this that nonlocal bias effects have a minimal impact on power spectrum predictions in the weakly nonlinear regime. This prediction is in good agreement with our estimates for nonlocal effects in the power spectrum from the sampling method below. This should be contrasted with Fig.~\ref{fig:Pk_NLcomponent}, from which one may conclude that the effect of the nonlocal bias on the power spectrum is quite significant. This is because local biasing prescription constructed at the scale $R$ yields a good effective description for the halo field and it can partly absorb the nonlocal effects. Note that without applying Eq.~(\ref{eq:kR}), the results hardly make any sense.


 \begin{figure}[!t]
\centering
 \includegraphics[ width=0.9\linewidth]{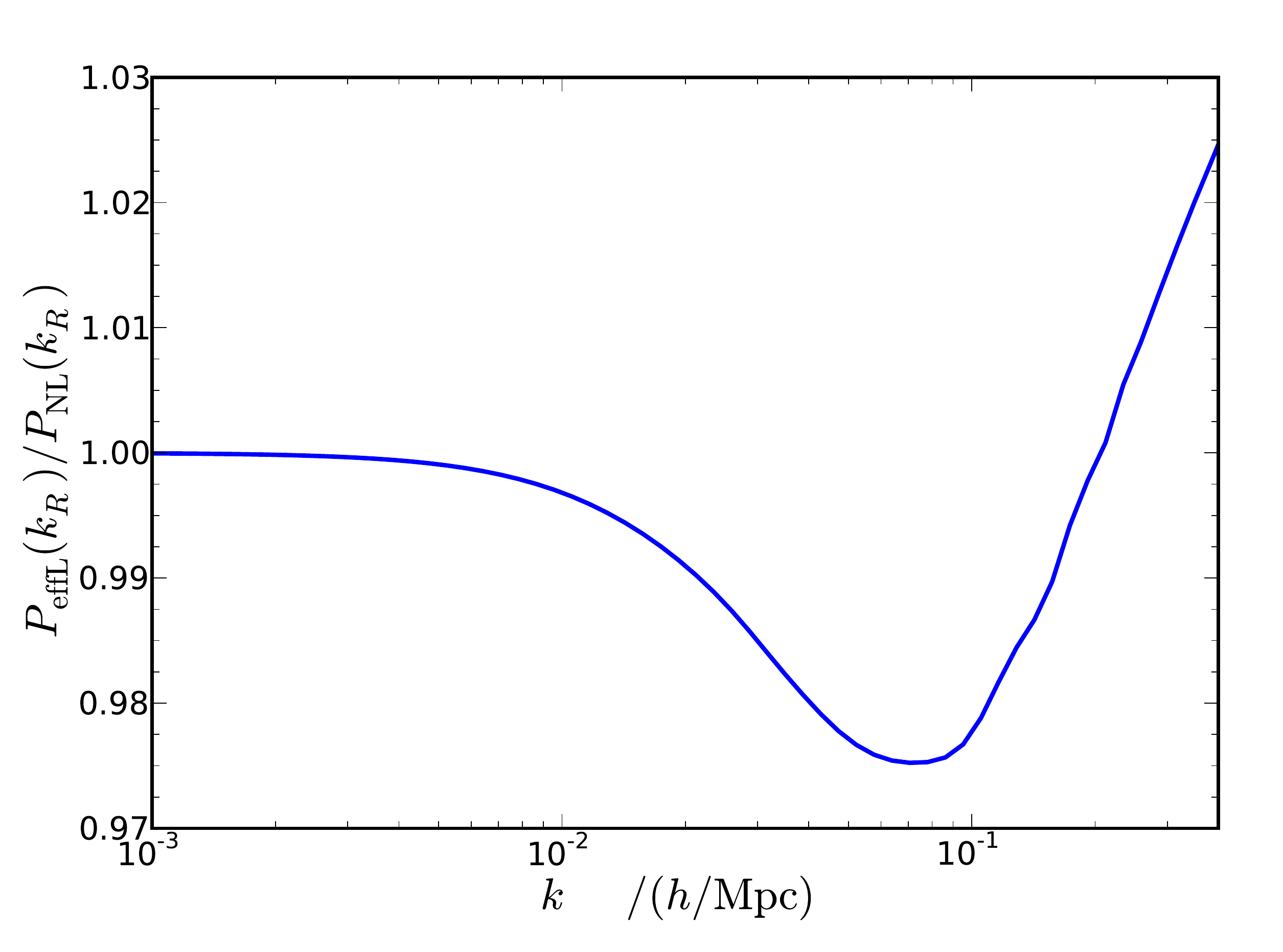}
 \caption{ The ratio of the effective local model power spectrum to the underlying nonlocal model. The power spectra are computed upto 1-loop, and the scale dependence of local biasing prescription has been taken into account.  Nonlocal bias effects have a small impact on the power spectrum.}
 \label{NlocTOloc} 
 \end{figure}

\subsection{ Errors from assuming local bias }
\label{sec:NonlocalvsLocal}

In this subsection, we shall approach the following question in more detail: suppose that the underlying biasing model is given by some nonlocal prescription, such as Eq.~(\ref{eq:NLbiasing}), how biased the results would be if the local biasing prescription is assumed?  To study this in a controlled manner, we construct what we call \textit{man-made} nonlocal halo density from the dark matter density field with some input bias parameters. We will compare the power spectrum obtained in four different ways: the numerical power spectrum computed directly from the man-made  halo density field, the 1-loop perturbation theory results from the input nonlocal model, the power spectrum from shuffling of the man-made halo density field, and the 1-loop results from the effective local model.  Our approach is similar to Ref.~\cite{Pollacketal}, in which artificial local models are constructed to test the importance of higher order terms in the bispectrum. 

As found in \cite{NLBias} from numerical simulations,  the dominant nonlocal contribution to biasing prescription originates from the $\mathcal{G}_2 $ term, and so for simplicity, we shall consider the nonlocal model up to second order in $\delta $
\begin{equation}
\label{eq:ManMadeNLModel}
\delta_{\mathrm{h} R} = b_1 \delta_{ R} + \frac{ b_2 }{ 2 }  \delta_{R}^2  + \gamma_2  \mathcal{G}_{2 R}.    
\end{equation}
We shall refer to this model as the  $\mathcal{G}_2  $ model.  The field $\delta_R$ is obtained by simply smoothing the nonlinear dark matter density $\delta $ from the $N$-body simulation. In Eq.~(\ref{eq:NLbiasing}), $\mathcal{G}_2 $ is constructed from the velocity potential. As obtaining the velocity potential is cumbersome and the linear power spectrum is close to the nonlinear velocity divergence power spectrum \cite{PueblasScoccimarroVel}, we shall use the linear gravitational potential instead of  the nonlinear velocity potential. The error committed here is third order in linear $\delta_{\rm L } $. To construct $\mathcal{G}_{2 R}$, we first  compute $\nabla_{ij} \Phi = ( \bold{k}_i   \bold{k}_j / k^2 ) \delta_{\rm L}( \bold{k} )$ in Fourier space. We then inverse Fourier transform to get  $\nabla_{ij} \Phi  (\mathbf{x})$, from which we construct $\mathcal{G}_2$ using $\mathcal{G}_2 = (\nabla_{ij} \Phi )^2  - ( \nabla^2 \Phi )^2  $.  Finally, after applying the smoothing window, we arrive at $\mathcal{G}_{2 R}$. 

\begin{figure*}[!t]
\centering
\includegraphics[ width=0.9\linewidth]{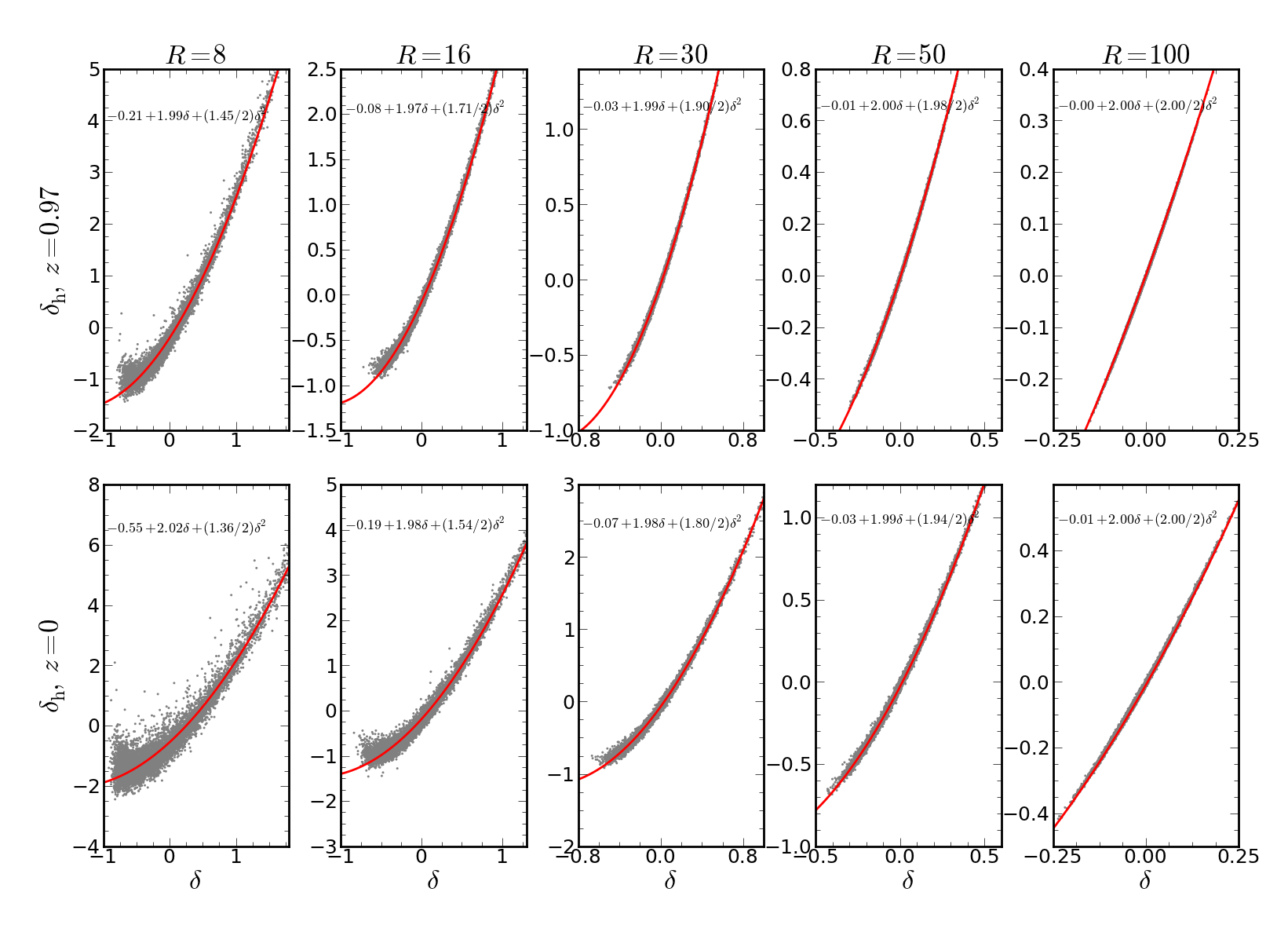}
\caption{ The scatter plots from a ``man-made'' halo density field constructed using $b_1=2$, $b_2=1.34$ and $\gamma_2=-0.5$. Top-hat window of smoothing scales 8, 16, 30, 50 and 100 Mpc/$h$ are used. Two redshifts are considered: 0.97 (the first row) and 0 (the second row).  Shown on top of the scatter plot is the best fit model using local bias. The best fit $b_1^R$ agrees with the input parameter for all the smoothing scales, while the best fit $b_2^R$ agrees with the naive estimated value 2.01 only for large smoothing scales.   }
\label{fig:scatterplot_800_c2-1_lindelta}
\end{figure*}

We will apply the shuffling technique to the man-made $\delta_{\mathrm{h} R} $. Although we anticipate this to be equivalent to the local biasing procedure, we will explicitly check this by fitting to the scatter plot obtained from Eq.~(\ref{eq:ManMadeNLModel}) with the local model
\begin{equation}
\label{eq:LFitModel}
\delta_{\mathrm{h} R} = b_0^R +   b_1^R \delta_{R} + \frac{ b_2^R }{ 2 } \delta_{R}^2. 
\end{equation}
We shall see that $b_1^R$ is largely unaffected, while $b_2^R$ will be systematically shifted due to the presence of $\mathcal{G}_{2 R}$.

Let us first estimate the value of $b_2^R$. The term $\mathcal{G}_2$ in real space is given by 
\begin{eqnarray}
\mathcal{G}_2 (\bold{x} )& = & \int \diff^3 q_1  \diff^3 q_2  \mathrm{e}^{i ( \bold{q}_1 +  \bold{q}_2 )  \cdot \bold{x} }  ( \mu_{12}^2 -1 )        \nonumber  \\
 &  \times  &  \delta_{\rm L}( \bold{q}_1  ) \delta_{\rm L} ( \bold{q}_2 ).      
\end{eqnarray}
Because of spatial homogeneity, we can consider $\mathcal{G}_2 (\bold{0})$ only. The average of the kernel, $ \mu_{12}^2 -1$, \textit{i.e.}~$\int \diff^3 q_1 \diff^3 q_2     ( \mu_{12}^2 -1 ) / ( \int \diff^3 q_1 \diff^3 q_2  )$, is $-2/3$. This approximation has been checked against simulations and found to be good for a smoothing scale of 40 Mpc/$h$ of a top-hat window \cite{NLBias}. Thus, approximating the kernel by its average, we expect to have 
\begin{equation}
\label{eq:AveLBias}
b_0^R=0, \quad   b_1^R=b_1, \quad  b_2^R = b_2 - \frac{4}{3 }  \gamma_2 .
\end{equation}
Note that we have ignored the impact of the smoothing window in this simple estimate. 

The shift of the local bias parameter $b_2$ depends on the form of the nonlocal term considered. The nonlocal term generated by gravitational evolution has the form of $\mathcal{G}_2 $, which is naturally fixed by Galilean invariance~\cite{NLBias}. However, one could also shuffle some of local term amplitude inside $\mathcal{G}_2 $ to the local quadratic bias and have a nonlocal operator proportional to the quadrupole, e.g. $( \nabla_{ij} \Phi )^2 - 1/3  ( \nabla^2 \Phi )^2 $ as in~\cite{McDonaldRoy}, in which case this nonlocal term does not shift the local quadratic bias parameter upon spherical averaging.

The input bias parameters are chosen to be $b_1=2$ and $b_2=1.34$, and four different values of $\gamma_2$ for comparison: $-0.5$, $-0.2$, $0.1$ and 1, although we will only show plots for $\gamma_2=-0.5$. In Fig.~\ref{fig:scatterplot_800_c2-1_lindelta}, we show the scatter plots for the man-made halo density field constructed with $\gamma_2= -0.5$, which is roughly the mean value of $\gamma_2 $ found in \cite{NLBias}. Top-hat windows with $R$ being 8, 16, 30, 50, and 100 Mpc/$h$ respectively are used. On top of the scatter plots, we have shown the best fit using the effective local model Eq.~(\ref{eq:LFitModel}). If our estimate Eq.~(\ref{eq:AveLBias}) were exact, we would expect $b_1^R = 2$ and, for $\gamma_2=-0.5$, $b_2^R $ to be 2.01. Indeed the best fit $b_1^R$ is close to 2 irrespective of the smoothing scale.  On the other hand, from Fig.~\ref{fig:scatterplot_800_c2-1_lindelta}, the $b_2^R $ estimate is accurate only for large smoothing window. As $R$ increases, the best fit $b_2^R  $ increases and tends to the estimated value. This behavior is most likely due to the effect of smoothing, note that in the man-made halo field there is only one window function acting on $\mathcal{G}_2 $, and we have converted the shift on the quadratic bias in Eq.~(\ref{eq:AveLBias}) ignoring this (i.e. $\delta_{R}^2$ contains two windows, not one). Also note that $b_0^R$ is non-zero for small $R$, this can be expected from doing a conditional average of $\mathcal{G}_2 $ at fixed $\delta $ (rather than our spherical, unconditional, average), which gives an extra contribution to Eq.~(\ref{eq:AveLBias}) of $b_0^R=-\sigma^2_R/3$.  We see a stronger amplitude than $-1/3$, and this may again related to the mismatch in the number of window functions in the conditional average argument here.   Note that the values of the bias parameters are not chosen carefully to ensure $\delta_{\mathrm{ h} R} \geq  -1$. However, for a reasonably large smoothing scale, the man-made density obeys $\delta_{\mathrm{ h } R  } \geq  -1$.

In this deterministic model, the scatter originates from the projection of the $\mathcal{G}_2 $ term onto the $\delta_{\rm h}-\delta $ plane. We can compare the scatter due to the projection effect of the nonlocal terms with that in the standard scatter plots from simulations. If we consider halos from simulations with values of  $b_1^R$ and $b_2^R$ similar to those we considered here, and with the same smoothing scale, even when we take some large value  such as $\gamma_2 =1 $, the scatter in the man-made scatter plot is still much smaller than that in the simulation  scatter plots. Thus the scatter in the scatter plots from simulations cannot be solely explained by the projection of the nonlocal terms alone, similar to what is found in \cite{NLBias}.  For other values of $\gamma_2$, the same trend in the scatter plots holds, and so we do not show them here.  Of course, the smaller the value of $| \gamma_2 |$, the smaller the scatter in the man-made scatter plot.   
 


\begin{figure*}[!t]
\centering
\includegraphics[ width=0.9\linewidth]{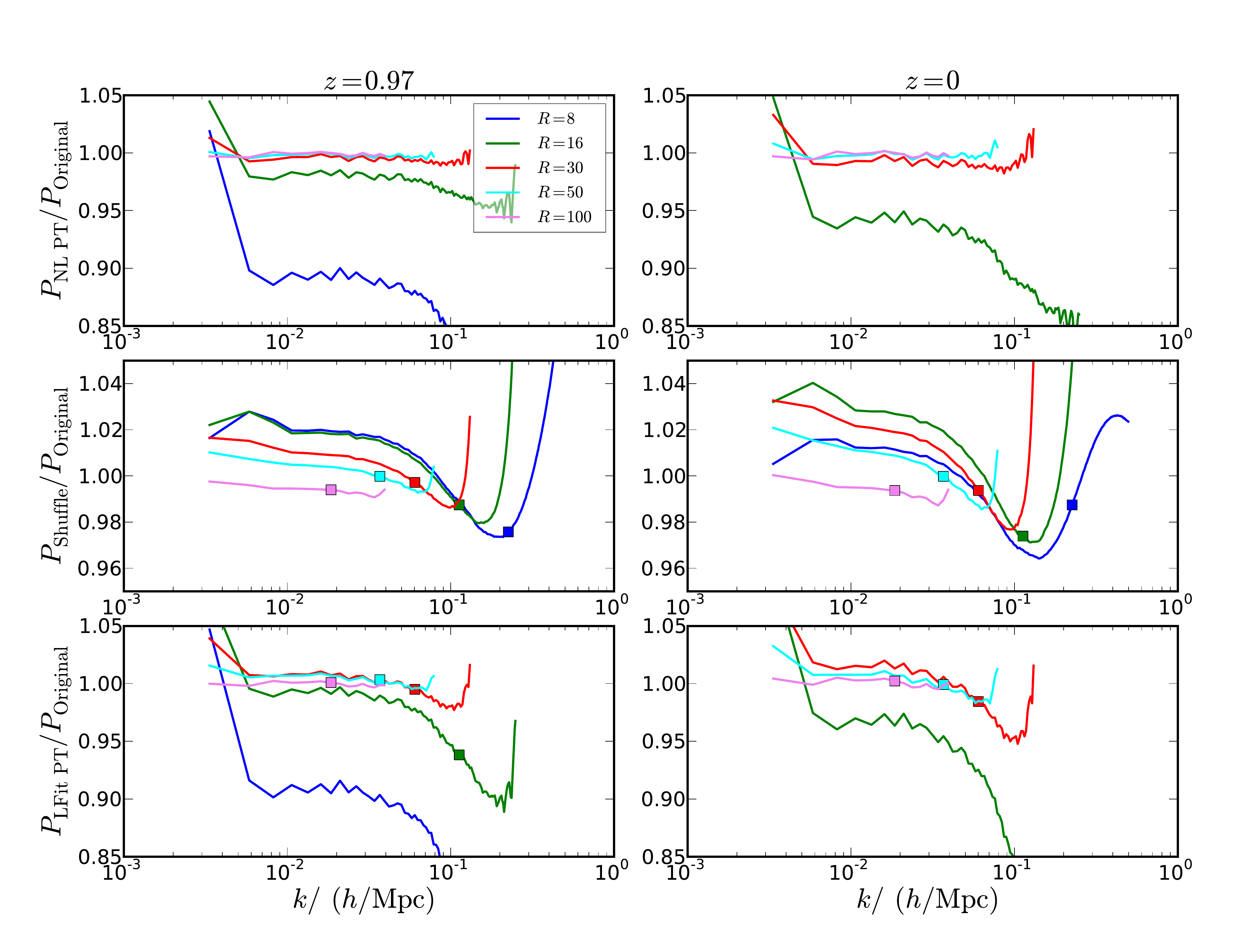}
\caption{ Different power spectrum calculations compared to the true (original) power spectrum of the ``man-made'' halo density field constructed with $\gamma_2 = -0.5 $ at various smoothing scales.  In the first row, we show the power spectra obtained from 1-loop perturbation theory using the nonlocal model supplemented with the input bias parameters. In the second row, we show the power spectra from shuffling of the ``man-made" halo field. Shown in the last row are the 1-loop perturbation theory results using the effective local model with the local bias parameters fitted from the scatter plots. All of them are normalized by the corresponding original power spectra.  Two redshifts are considered: 0.97 (left column) and 0 (right column). In each plot, five smoothing scales are used: 8, 16, 30, 50 and 100 Mpc/$h$ (same as Fig.~\ref{fig:scatterplot_800_c2-1_lindelta}).  Some of curves with small smoothing scales fall outside the range shown. Also, for each curve we show only up to the scale at which it starts to rapidly oscillate due to the top-hat window function.  For the shuffling and the effective local model results, we have also plotted the points obtained using Eq.~(\ref{eq:kR}) (squares).          }
\label{fig:Pk_ManMade_Shuffle_NL_LFit_WTH_c2-1}
\end{figure*}

In Fig.~\ref{fig:Pk_ManMade_Shuffle_NL_LFit_WTH_c2-1},  we compare the power spectrum obtained from perturbation theory using the nonlocal model, shuffling, and perturbation theory using the effective local model against the numerical power spectrum from the man-made halo density. In  Fig.~\ref{fig:Pk_ManMade_Shuffle_NL_LFit_WTH_c2-1}, $\gamma_2$ is again chosen to be $-0.5$.  
The nonlocal one-loop perturbation results include the local contribution  Eqs.~(\ref{eq:Ph1_1}--\ref{eq:Ph11_11}) and the nonlocal contribution Eqs.~(\ref{eq:Pb1gamma2_2}--\ref{eq:Pgamma2gamma2}), computed using the input bias parameters. We note that Eq.~(\ref{eq:Pb1gamma2_1}) is not present because we have used the linear potential to construct $\mathcal{G}_2  $ instead of the nonlinear velocity potential. The one-loop effective local model calculation includes only the local contribution Eq.~(\ref{eq:Ph1_1}--\ref{eq:Ph11_11}), but the bias parameters are the best fits in the scatter plots. 

The nonlocal bias perturbative results agree very well with the original man-made model results for large scales (the first row in Fig.~\ref{fig:Pk_ManMade_Shuffle_NL_LFit_WTH_c2-1}). However, as the smoothing scales decreases, the agreement deteriorates as the smoothing scale decreases, especially for 16 and 8 Mpc/$h$. The agreement is also better for $z=0.97$ than for $z=0$.  This is consistent with larger nonlinearity in the constructed field when the smoothing scale is small or the redshift is low, and so perturbation theory is not so accurate.

After shuffling (middle panels in Fig.~\ref{fig:Pk_ManMade_Shuffle_NL_LFit_WTH_c2-1}), the results are indeed biased, especially for small smoothing scales, and the perturbation results using the effective local model (bottom panels) share qualitatively similar trends as the shuffling results although sometimes it is not clear-cut. Let us first apply the insights from the local model, especially the behavior of the effective bias parameters, to interpret the shuffling results. For the smoothing scale as large as 100 Mpc/$h$, the loop correction is negligible, and so its value is mainly determined by the best fit $b_1^R $.  As the smoothing scale decreases, the loop corrections become more important. As we saw in Section \ref{sec:NLPk}, the corrections due to the nonlocal term are significant only for $k \gtrsim 0.1 \, h/\mathrm{Mpc} $, while the local correction, especially Eq.~(\ref{eq:Ph1_12}), which is approximately proportional to $P_{\rm L}(k)$, is already significant at large scales. As a result of a systematically enhanced $b_2^R$ due to $\mathcal{G}_2 $, the shuffling power spectrum is higher than the original one for $R \leq 50 \, \mathrm{Mpc}/h $. However, we see that in Fig.~\ref{fig:scatterplot_800_c2-1_lindelta} for small $R$ such as 16 and 8 Mpc/$h$ as $R$ decreases the effective $b_2^R $ also decreases, this effect is especially significant at $z=0$. This reduction in  $b_2^R $ for small $R$ can explain why the shuffled power spectrum at $z=0$ suddenly plummets for $R=8 \, \mathrm{Mpc}/h $. The 1-loop results based on the effective local model are consistent with the trends  discussed for the shuffled halo power spectrum, but generally the effects are stronger. Again perturbation theory fairs badly for fields constructed using small smoothing scales or at low redshift.  In some sense, shuffling is the best local biasing prescription can achieve because all the information in the whole scatter plot is used, and the power spectrum is computed numerically.

We would like to clarify the relation between the man-made halo density field here and the halo field from simulations.  The man-made halo density are constructed at some scale $R$ with the input bias parameters, while we have seen that scatter bias parameters of the halo density field run with scale. Hence we can regard the man-made density field as a ``snapshot'' of the halo density from simulation at scale $R$. This viewpoint is in line with our formulation of local bias. 

The nonlocal term $\mathcal{G}_2 $ with negative $\gamma_2 $  after the shuffling increases the effective $b_2^{ R } $, and so the shuffling power spectrum is higher than the original model at large scales. This agrees with the trends that we saw in Section \ref{sec:numerical_results}. However, besides the contribution arising from $\mathcal{G}_2  $, there is another piece from the local $b_2$ (in the  spherical approximation it is given by Eq.~\ref{eq:AveLBias}).  Because it is difficult to separate these two components, inspecting $P(k)$ at large scales is not an effective way to detect the nonlocal bias. In fact, we argued in Section \ref{sec:PkWithBiasRunning} that the biasing prescription obtained with smoothing scale $R$, the effective bias parameters can absorb some of the effects due the nonlocal terms at that scale.  In the case of bispectrum, which will be considered elsewhere, is expected to be more effecitve in detecting the nonlocality in the biasing prescription.

So far we have discussed the behavior of these models in the small $k$ region, $ k \lesssim 0.05 $ $h$/Mpc. According to our formulation, for the halo biasing model constructed at scale $R$, we should focus on the $ k\sim 1/R $ only. 

For each curve (second and third rows in Fig.~\ref{fig:Pk_ManMade_Shuffle_NL_LFit_WTH_c2-1}), we have also shown the point  mapped using Eq.~(\ref{eq:kR}). Overall for the five smoothing scales considered, the deviations of the shuffling results from the input nonlocal model are within 3\% up to $k\sim 0.2$ $h$/Mpc. The difference is small, but is still slightly larger than what we found between the shuffling results and the original simulations. However, we only include the $\mathcal{G}_2$ term using the linear potential, in particular the difference between linear potential and the nonlinear potential may not be negligible in the nonlinear regime. Note that Fig.~\ref{NlocTOloc} is quite similar to curves formed by the square symbols in Fig.~\ref{fig:Pk_ManMade_Shuffle_NL_LFit_WTH_c2-1}, especially the middle panel. Thus both approaches suggest that the effect of the nonlocal bias on the power spectrum is with a few per cents up to the weakly nonlinear regime.  Overall, the good agreement supports our claim that the effective description constructed at the scale $R$ is accurate at that scale alone.

\section{Conclusions}
\label{sec:Conclusion}

In order to test the validity of the local biasing prescription and to examine the impact of the nonlocality,  we constructed a halo field by sampling the dark matter-halo distribution from simulations. The method is implemented by first constructing a scatter plot between $\delta_R $ and $\delta_{ \mathrm{h} R } $, and then dividing $ \delta_R $ into bins and shuffling the halos $\delta_{\mathrm{h}R } $  within the same dark matter density bin. Because we  randomly permute the halos within the same $\delta_R$ bin, the halo field constructed in this way has the same scatter plot as the original one. This method can also be reduced to a mean (scatterless) biasing prescription by replacing the halo density by its mean in the bin. We then compute numerically the power spectrum of the halo fluctuations defined in this way. However, we find that although the shuffling method keeps the (one-point) scatter in the scatter plot, any stochastic noise (including shot noise) is destroyed by the shuffling process at the two-point level, in the sense that the noise is eliminated by ensemble average when computing the power spectrum (see Appendix). Thus we obtain that the difference between the shuffling and the mean methods is negligible at the end.

The method developed here differs from the usual test of local biasing in several aspects. First the shuffling method keeps all the information in the scatter plot between $\delta_R$ and $\delta_{\mathrm{ h} R } $. As we do not fit to the distribution by a polynomial, we essentially test the series in Eq.~(\ref{eq:EulerianLocalBiasing}) without truncation. Often, perturbation theory is invoked to compute the correlators in testing local biasing. When such predictions break down, it is not clear if local biasing fails and/or perturbation theory breaks down. As we compute the power spectrum numerically, \textit{i.e.} nonperturbatively, we can test the validity of local bias without worrying about the breakdown of perturbation theory. More importantly, this enables us to extend the test towards the nonlinear regime. 

In what has become to a large extent the standard interpretation of local bias in the literature~\cite{Heavens1998,Smith2007,McDonaldReBias,RothPorciani,JimenezDurrer,WangSzalay}, loop corrections from nonlinear bias generate significant corrections at large scales; for the power spectrum this is essentially a renormalization of the linear bias and a renormalization of the shot-noise. Depending on the particular implementation, such renormalizations are poorly determined (depending on an unknown smoothing scale $R$) or arbitrarily left as free parameters. This standard perturbative calculation is illustrated in Fig.~\ref{fig:bc_PT2}, showing the dependence on $R$. Given our sampling method, we can perform a similar calculation (see Figs.~\ref{fig:bc_Sim_Shuffle_Mean_800} and~\ref{fig:bh_Sim_Shuffle_Mean_800}), again at fixed $R$, and check that the non-perturbative results are quite similar to those obtained perturbatively. Thus, the problem originates from the interpretation of local biasing itself, not in perturbation theory. 

Our proposal is that local bias parameters obtained using a smoothing scale $R$ can only be applied to computing the halo power spectrum at scales $k\sim 1/R$ (or, more precisely, Eq.~\ref{eq:kR}). Our calculations thus can automatically include the running of bias parameters (see Fig.~\ref{fig:running_bias}) and give vanishing small loop corrections at low-$k$ since the smoothing scale $R\to \infty$ as $k\to 0$.  This brings back a fairly good agreement at large scales when bias parameters measured from a scatter plot are used and the calculation done perturbatively (see Fig.~\ref{fig:bc_SimShuffleMean_TreeLoop2_3}). For example, we obtain 4 \% agreement up to  $k\sim0.1$ $h$/Mpc at $z=0.97$ and 7 \% agreement up to $k\sim0.1$ $h$/Mpc at $z=0$. A much better agreement is obtained  when the calculation is done non-perturbatively via our sampling method. We stress that our interpretation of the smoothing scale in local bias is precisely the same as originally put forward by Fry \& Gazta\~naga \cite{FryGaztanaga} for the computation of counts in cells. In that case there is  no contribution of non-linear bias loops to the variance of counts in cells in the large-cell limit (linear bias and shot-noise renormalizations vanish).  

Under the ``standard interpretation", nonlinear bias loops generate a constant power at very large scales (shot-noise renormalization), which in principle can overwhelm the primordial scale-invariant spectrum that drops with $k$ as $k \to 0$. By contrast, our interpretation posits that all nonlinear bias loops are highly suppressed at low-$k$ as the loop momenta only run up to $\simeq k$. We looked for evidence of shot noise renormalization at wavenumbers below the turn-over of the power spectrum $k<0.01~h$/Mpc and have not found any, except for the sub-Poisson effect due to halo exclusion (see Figure~\ref{fig:b1_LargeScaleFit_MW}). This supports our proposal for how nonlinear bias loops should be treated, and that the halo power spectrum in the low-$k$ limit is  given by linear bias with no extra contributions (apart from non-perturbative corrections such as the shot noise and halo exclusion). We also find from the sampling method that $b_{\rm h}$ and $b_{\rm c}$ agree with each other at large scales, which lends further support to our approach.

To help interpret the sampling results, we also constructed and artificial ``man-made" halo density using the nonlocal bias model described in \cite{NLBias}, and apply the sampling technique to it. The middle panel in Fig.~\ref{fig:Pk_ManMade_Shuffle_NL_LFit_WTH_c2-1} shows that, after the mapping given by Eq.~(\ref{eq:kR}) denoted by square symbols, the results from shuffling (which makes the bias local) are very close to the results directly from the input model. This indicates that the nonlocality of bias does affect the power spectrum only very slightly, at the few percent level. This is also in agreement with our estimates from Fig.~\ref{NlocTOloc} using perturbation theory.

\acknowledgements 
We thank Tobias Baldauf, Vincent Desjacques, Aseem Paranjape, Cristiano Porciani, Fabian Schmidt, Uros Seljak, Ravi Sheth, and Robert Smith for useful discussions. We owe additional thanks to Ravi Sheth for suggesting the mass-weighting test done in Figure~\ref{fig:b1_LargeScaleFit_MW}.  The simulations presented here are part of the LasDamas collaboration suite\footnote{\tt http://lss.phy.vanderbilt.edu/lasdamas}  and were run thanks to a Teragrid allocation and the use of RPI and NYU computing resources.  This work was partially supported by grants NSF AST-1109432 and NASA NNA10A171G. KCC acknowledges the support of the Mark Leslie Graduate Assistantship.

\appendix

\section{More on Poisson shot noise and stochastic noise}
\label{sec:appendix}
\subsection{ Poisson shot noise }
\label{sec:PoissonShotNoise}

To derive the standard Poisson shot noise result, the density field is represented by a distribution of point particles. Then the number density is given by
\begin{equation}
n_0 (\mb{x} ) = \sum_i \Ddel ( \mb{x} - \mb{x}_i ). 
\end{equation}
The reason for the subscript 0 will become clear in a moment. The covariance matrix  of the number density field  is given by 
\begin{equation}
\label{eq:nn_correlation}
\langle n_0  (\mb{x}_1 ) n_0 (\mb{x}_2 )  \rangle = \sum_{i,j} \langle \Ddel ( \mb{x}_1 - \mb{x}_i ) \Ddel ( \mb{x}_2 - \mb{x}_j ) \rangle. 
\end{equation}
Because Dirac $\delta$ functions are very singular when they step on each other, special care is taken for $i=j$.  The sum on the right hand side of Eq.~(\ref{eq:nn_correlation}) is split into two parts, one with $i=j$ and another with $i \neq j$. Therefore, Eq.~(\ref{eq:nn_correlation}) can be written as 
\begin{equation}
\langle n_0  (\mb{x}_1 ) n_0  (\mb{x}_2 )  \rangle = \bar{n}_0 \Ddel( \mb{x}_1 - \mb{x}_2 ) + \bar{n}^2_0 [ 1 + \xi_{ \rm c} (x_{12} ) ],
\end{equation}
where $\bar{n}_0 $ is the mean number density 
\begin{equation} 
\bar{n}_0 =   \sum_i  \langle \Ddel ( \mb{x} - \mb{x}_i ) \rangle, 
\end{equation}
and we have replaced the $i \neq j $ part with the continuous correlation function $\xi_{\rm c}$ as
\begin{equation}
 \sum_{  i \neq  j     } \langle \Ddel ( \mb{x}_1 - \mb{x}_i ) \Ddel ( \mb{x}_2 - \mb{x}_j ) \rangle = \bar{n}_0^2 [ 1 + \xi_{\rm c}( x_{12} )  ]. 
\end{equation}
Writing $\langle n_0(\mb{x}_1 ) n_0(\mb{x}_2 )  \rangle$ as $\bar{n}_0^2 [ 1 + \xi( x_{12} )]$ we arrive at 
\begin{equation}
\xi( x_{12} ) = \xi_{\rm c} (x_{12}) + \frac{1}{ \bar{n}_0 } \Ddel ( \mb{x}_1 -\mb{x}_2 ). 
\end{equation}
Upon Fourier transform, we get the familiar result 
\begin{equation}
\label{eq:constantweight}
P(k) = P_{\rm c}(k)  + \frac{1}{ ( 2 \pi )^3 } \frac{ 1 }{ \bar{n}_0 }.
\end{equation}
This derivation here highlights that the Poisson shot noise arises as long as the objects are point particles. In simulations, the dark matter particles and halos are indeed treated as point particles.

Now we consider the weighted number density 
\begin{equation}
n_{w} ( \mb{x} ) = \sum_i  w_i \Ddel ( \mb{x} - \mb{x}_i ) ,
\end{equation}
where $w_i $ is some weight. A similar exercise yields
\begin{equation}
\xi ( x_{12} ) = \xi_{\rm c} (x_{12}) + \frac{ \overline{ n_w^2 } }{ \bar{n}_w^2 } \Ddel ( \mb{x}_1 -\mb{x}_2 ), 
\end{equation} 
where $\bar{ n }_w$ and  $ \overline{n_w^2} $ are given by 
\begin{eqnarray}
\bar{ n }_w     &=& \sum_i  \langle   w_i  \Ddel (  \mb{x}  -  \mb{x}_i   )  \rangle  , \\
\overline{ n_w^2  } &=& \sum_i  \langle   w_i^2  \Ddel (  \mb{x}  -  \mb{x}_i   )  \rangle  .
\end{eqnarray}
Therefore the Poisson shot noise is now given by 
\begin{equation}
\label{eq:massweight}
P (k ) = P_{\rm c} ( k ) +  \frac{1}{ ( 2 \pi )^3 }  \frac{ \overline{ n_w^2 } }{ \bar{n}_w^2 }. 
\end{equation} 

\begin{figure*}[!htb]
\centering
\includegraphics[ width=0.9\linewidth]{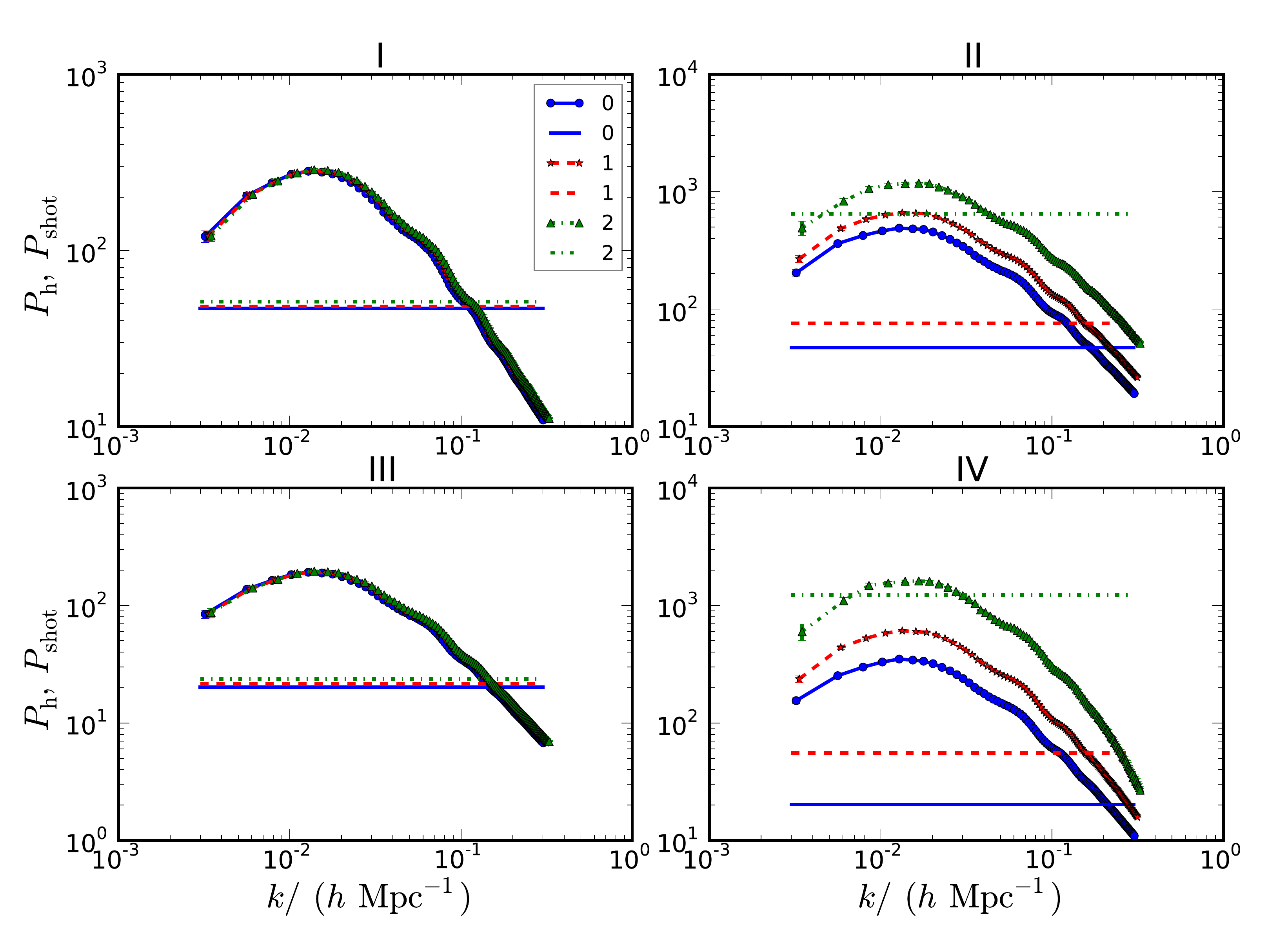}
\caption{ The halo power spectrum with shot noise subtracted (lines with symbols) and the corresponding shot noise (straight lines). Mass weighting with weight $m^k$ with $k=0$(blue), 1(red), and 2(green) are shown for each halo group. Mass weighting has little effect for the low mass groups (I and III), while the power spectrum and its shot noise are significantly increased for the massive groups (II and IV).  }
\label{fig:Ph_Pshot}
\end{figure*}

When the weight is constant, we recover Eq.~(\ref{eq:constantweight}). One physically interesting weighting scheme is weighting by mass or even some power of it \cite{SeljakHamausetal_2009, HamausSeljaketal_2010, CaiBernsteinSheth2011}  Mass weighting would allow us to select the more biased objects in the sample.  One can show that for the Poisson model, the shot noise is minimized for equal weight. However, the halo power spectrum also increases as the more biased halos gain weight. Nonetheless we find that the increase in shot noise level is generally larger than the enhancement due to the halo bias. Ref.~\cite{SeljakHamausetal_2009, HamausSeljaketal_2010} reported mass weighting can reduce shot noise because the sub-Poisson effects due to halo exclusion become more significant.  
In Fig.~\ref{fig:Ph_Pshot} we show the halo auto power spectrum with shot noise subtracted and the corresponding shot noise for various halo groups. Mass weighting $m^k$, with $k=0$, 1, and 2 are considered. For the low mass groups (I and II) mass weighting has little effects on the overall power spectrum and its shot noise as the mass bins are quite narrow. On the other hand, for the high mass groups (II and IV), mass weighting increases the overall power spectrum and the shot noise is enhanced significantly. For the weighting by $m^2$, the shot noise is comparable or higher than the signal on scales $k \lesssim 0.1 \, h/\mathrm{Mpc} $. 

\subsection{Smoothing and shuffling of the stochastic noise } 
\label{sec:ShufflingStochasticNoise}

We first discuss the case when the density fields are smoothed with some window functions. Here we denote the smoothing operator as $\hat{R}$ (this is the usual convolution with a top-hat window function in real space). Although after smoothing the particles spread out, the smoothing operator is constant with respect to the ensemble average operator, \textit{i.e.}~the smoothing operator commutes with the ensemble average operator.  The smoothing operator leaves the original field coherent, and so we have
\begin{equation}
\langle  \hat{R}_1 n( \mb{x}_1 ) \hat{R}_2 n( \mb{x}_2 )   \rangle =  \hat{R}_1 \hat{R}_2 \langle  n( \mb{x}_1 ) n( \mb{x}_2 ) \rangle.  
\end{equation}
Therefore when the fields are smoothed by some window function the shot noise still contributes to the power spectrum except with the factors of the window functions (Eq.~(\ref{eq:Pkshot})).

In addition to the mean biasing relation, we now include a stochastic noise $\epsilon_R $, then $\delta_{\mathrm{ h}R } $ is given by 
\begin{equation}
\label{eq:LocalBiasingNoise}
\delta_{\mathrm{ h} R} = f_R( \delta_R ) + \epsilon_R ,
\end{equation}
with 
\begin{eqnarray}
f_R( \delta_R ) = \langle \delta_{\mathrm{h} R} | \delta_R   \rangle,    \\
\epsilon_R =  \delta_{\mathrm{h} R} - f_R( \delta_R ) .
\end{eqnarray}
To be general, $\epsilon_R $ can include sources of stochasticity other than the Poisson shot noise. The stochastic random field $\epsilon_R $ satisfies  
\begin{equation}
\langle  \epsilon_R \rangle = 0 .
\end{equation}
We do not assume the two-point correlation function of $\epsilon_R$  to be Poisson, but we assume it to be uncorrelated with the dark matter density $\delta_R  $. 

Eq.~(\ref{eq:LocalBiasingNoise}) consists of two parts, the coherent part ($f_R( \delta_R )$), whose coherent length is large or comparable to the smoothing scale, and the incoherent part ($ \epsilon_R $), whose coherent length is much smaller than the smoothing scale. These two parts respond to smoothing differently. The coherent part is affected little by smoothing, while the incoherent part is greatly suppressed.

When we do shuffling of the halo overdensity within the same bin of $\delta_R $, we essentially apply some random distribution function to the density field. For convenience, we represent this process by the \textit{shuffling operator} $\hat{S}$. Note that this shuffling operator is a stochastic operator, and it does not commutes with the ensemble average operator. Thus the expectation value of the shuffled halo field is given by 
\begin{eqnarray} 
&  & \langle \hat{S} [  f_R ( \delta_R ( \mb{x}_1 )) + \epsilon_R ( \mb{x}_1 )  ]   \hat{S} [  f_R ( \delta_R ( \mb{x}_2 ) )  + \epsilon_R ( \mb{x}_2 )  ]    \rangle   \nonumber   \\
& = &  \langle f_R ( \delta_R ( \mb{x}_1 )) f_R( \delta_R ( \mb{x}_2 )) \rangle   + 
    \langle    \hat{S}  \epsilon_R ( \mb{x}_1 )  \hat{S}  \epsilon_R ( \mb{x}_2 ) \rangle  \nonumber  \\
\end{eqnarray}
We have used the fact that $\hat{S}$ is designed to leave $ f_R ( \delta_R ( \mb{x} )) $ invariant. On the other hand, we can think of $\hat{S}$ as a random distribution function
which distributes the ``parts'' of the particle everywhere, and so we can expect its contribution is negligible. Indeed this is consistent with the what we observe: difference between shuffling and sampling by the mean is negligible.

\begin{figure*}[!htb]
\centering
\includegraphics[ width=0.9\linewidth]{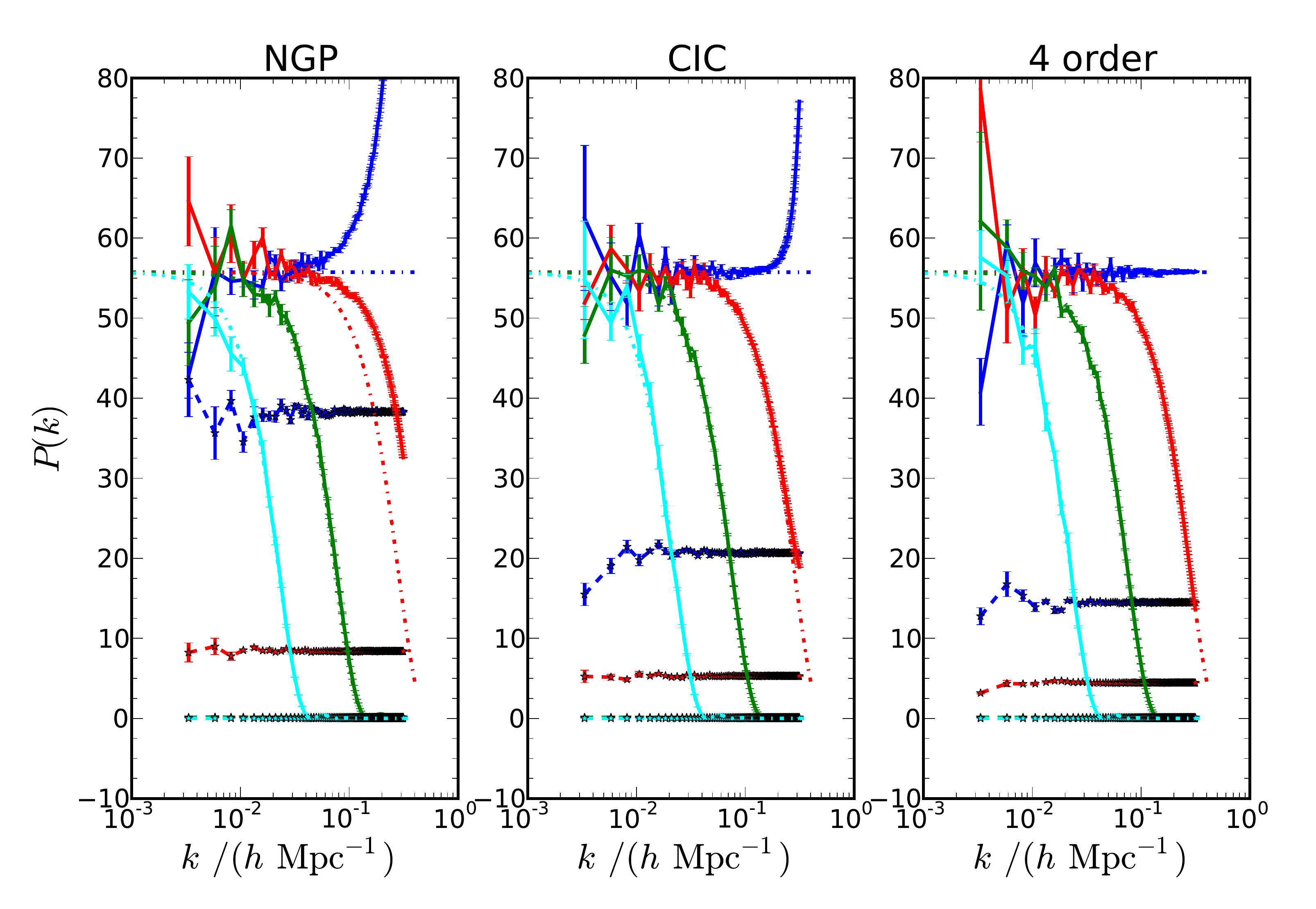}
\caption{    The original white noise power specturm and the one with shuffling.  The three columns correspond to  different interpolating kernels: NGP, CIC and $ 4^{\rm th} $ order (from left to right).  Within each subplot, we show the power specturm obtained with different smoothing scales  0 (blue), 8 (red),  30 (green) and 100 Mpc/$h$ (cyan) respectively. The original power spectrum is shown as data points with error bars connected with solid lines and the shuffled one as data points with error bars connected with dashed lines. Besides, we also show the expected white noise power spectrum using Eq.~(\ref{eq:Pkshot})  (dotted-dashed lines).    }
\label{fig:ShuffleWhiteNoise_NGP_CIC_4Order}
\end{figure*}

We did a simple test for the case of Poisson white noise. We generate $10^6  $ particles in a cubic box of  size 2400 Mpc/$h$ randomly. The particles are interpolated to a $240^3$ grid. We have considered three interpolating kernel with increasing smothness: NGP, CIC and a $4^{\rm th}  $ order kernel. As the order of the interpolation increases, each particle mass is interpolated to increasing number of neighbouring grid points, and so the resultant density field is more and more nonlocal. We then divide by the corresponding window function in Fourier space to sharpen the amplitude before computing the power specturm. For more details on interpolating kernels, see \cite{HockneyEastwood}. In the main text, only the $4^{\rm th}  $ order kernel is used.  We will apply the top-hat window functions of radius 0 (no smoothing), 8, 30, and 100 Mpc/$h$ respectively.

  We will compare the original power spectrum with various smoothing scales against the one obtained by shuffling. The shuffling preocedure is implemented as follows. After interpolating the particles to the grid points and smoothing the field with the window function in Fourier space, we inverse Fourier transform the field back to real space. The grids points are randomly partitioned into several parts, \textit{e.g.}~4, each part has equal number of particles.  Becasue of the randomness of the partition of the grid points, the number of parts is irrelevant. For example, we have checked that using 4 parts gives the same results as 24 parts. This is consistent with shuffling procedure used in the main text because we assume the patition of the grid points is based on the local values of $\delta_R  $, which is uncorrelated with the noise. The grid points are randomly permuted within the same part. The permuted grid points are used to compute the power specturm as usual, but without further applying a smoothing window.

In Fig.~\ref{fig:ShuffleWhiteNoise_NGP_CIC_4Order}, we show the original white noise power spectrum with various smoothing scales, and the corresponding power specturm obtained by shuffling. The original smoothed power spectrum (solid)  agrees with the expected power Eq.~(\ref{eq:Pkshot}) (dotted-dashed) rather well. The discrepancy we see for the low order kernels, especially NGP,  at medium to high $k$ regime is because dividing by the Fourier transform of the window function does not work so well for low order interpolating kernels. 

We now look at the shuffled power spectrum (dashed).  Although the correlation function of the original Poisson white noise is strictly local, both the smoothing process and the interpolation make the density field nonlocal. Random distribution of the density field values at the grid point in shuffling will destroy the correlation in the nonlocal density field. For the same interpolating function, the larger the smoothing scale, the greater the reduction in power after shuffling. Similarly the higher the order of the interpolating kernel, the lower the power after shuffling. 

In summary, in the numerical power spectrum computed in the usual way, the small scale stochasticity is preserved, while the stochastic noise is destroyed by the sampling method. 

\bibliography{sample_halo_final}

\end{document}